\documentclass[12pt,english]{article}
\usepackage[T1]{fontenc}
\usepackage[utf8]{inputenc}
\usepackage[pdftex]{geometry}
\usepackage{bm}
\usepackage{amsmath}
\usepackage{amssymb}
\usepackage[pdftex]{graphicx}
\usepackage{setspace}
\makeatletter

\providecommand{\tabularnewline}{\\}

\numberwithin{equation}{section}
\newcommand{\lyxaddress}[1]{
	\par {\raggedright #1
	\vspace{1.4em}
	\noindent\par}
}

\allowdisplaybreaks[1]

\usepackage{braketmod}

\usepackage{color}
\usepackage[pdftex]{hyperref}
\hypersetup{
colorlinks,
linktoc=page,
citecolor=blue,
linkcolor=blue,
urlcolor=blue
}

\date{}

\makeatother

\usepackage{babel}
\usepackage[style=phys,biblabel=brackets, chaptertitle=false, pageranges=false, eprint=true]{biblatex}
\begin{document}
\noindent\begin{minipage}[t]{1\columnwidth}%
\title{\textbf{Duality between $U\times U$ and $O\times USp$ theories}\\
\textbf{from $A_{3}=D_{3}$}}
\author{Naotaka Kubo\,\footnotemark}
\maketitle

\lyxaddress{\begin{center}
\vspace{-18bp}
$^{*}$\,\textit{Center for Joint Quantum Studies and Department of Physics,}\\
\textit{School of Science, Tianjin University, 135 Yaguan Road, Tianjin 300350, China}\vspace{-10bp}
\par\end{center}}
\begin{abstract}
We study the duality between the ABJ(M) theory at Chern-Simons level $k=4$ and the orientifold ABJ theory at Chern-Simons level $k=1$ by using the $S^{3}$ partition function. The partition function can be computed using the supersymmetric localization in terms of a matrix model, and we derive an ideal Fermi gas system by applying the Fermi gas formalism to the matrix model. By using this formalism, we show that the matrix model can be embedded in the partition function of $\hat{A}_{3}$ or $\hat{D}_{3}$ quiver theories, and the equality of these quiver theories, which comes from $\hat{A}_{3}=\hat{D}_{3}$, leads to exact relations of the matrix models. These exact relations can be used to check the identification of the flavor symmetries. The perfect agreement between the matrix models also implies that no decoupled sector is needed.
\end{abstract}
\end{minipage}

\renewcommand{\thefootnote}{\fnsymbol{footnote}}
\footnotetext[1]{\textsf{naotaka.kubo@yukawa.kyoto-u.ac.jp}}
\renewcommand{\thefootnote}{\arabic{footnote}}\medskip{}
\thispagestyle{empty}

\newpage{}

\setcounter{page}{1}

\global\long\def\bra#1{\Bra{#1}}%

\global\long\def\bbra#1{\Bbra{#1}}%

\global\long\def\ket#1{\Ket{#1}}%

\global\long\def\kket#1{\Kket{#1}}%

\global\long\def\braket#1{\Braket{#1}}%

\global\long\def\bbraket#1{\Bbraket{#1}}%

\global\long\def\brakket#1{\Brakket{#1}}%

\global\long\def\bbrakket#1{\Bbrakket{#1}}%

\tableofcontents{}

\section{Introduction\label{sec:Introduction}}

Supersymmetric localization technique \cite{Pestun:2007rz} is a powerful tool to obtain the exact results for supersymmetric theories. Typically, this technique provides an exact expression of a supersymmetric observable as a finite dimensional integral, which is called matrix model. Although this technique is first introduced for 4d theories, now it is applied for various dimensions, manifolds and observables (see \cite{Pestun:2016zxk} for a review). For example, the supersymmetric localization technique has been applied to 3d partition functions on $S^{3}$ \cite{Kapustin:2009kz} or on $S^{2}\times S^{1}$ (superconformal index) \cite{Bhattacharya:2008zy,Bhattacharya:2008bja,Kim:2009wb,Imamura:2011su,Kapustin:2011jm,Dimofte:2011py}. Although originally the partition function is defined by the path integral, after applying the supersymmetric localization, it reduces to a matrix model. One important application is for dualities. One can check dualities by showing the exact equality of observables between dual theories. However, the importance of this approach is not only for the check but also for extracting details of dualities. For example, one can find the correspondence of global symmetries by turning on corresponding deformation parameters like mass deformations and FI deformations. Another example is finding a decoupled sector, which was actually performed by using the $S^{3}$ partition function in, for example, the ``duality appetizer'' \cite{Jafferis:2011ns}.

In this paper, we study 3d $\mathcal{N}=5,6$ superconformal Chern-Simons theories, known as the ABJ(M) theories \cite{Aharony:2008ug,Aharony:2008gk,Hosomichi:2008jb}. They describe the low energy dynamics of M2-branes probing $\mathbb{C}^{4}/\Gamma$. Here, $\Gamma$ determines the gauge group. When $\Gamma=\mathbb{Z}_{k}$, the gauge group is ${\rm U}\left(N_{1}\right)_{k}\times{\rm U}\left(N_{2}\right)_{-k}$. The orientifold version of this theory is also known, whose gauge group is ${\rm O}\left(N_{1}\right)_{2k}\times{\rm USp}\left(2N_{2}\right)_{-k}$. In this case, $\Gamma=\mathbb{D}_{k}$, where $\mathbb{D}_{k}$ is the binary dihedral group of $4k$ elements. Because these theories are closely related to the string theory and M-theory, dualities between these theories have string/M-theoretical interpretation. For example, the Seiberg-like duality can be generated via the Hanany-Witten transition in type IIB string theory \cite{Giveon:2008zn}. 

There are other types of dualities which come from the M-theoretical geometry. As we discussed above, the ${\rm U}\times{\rm U}$ theory comes form the geometry of $\mathbb{C}^{4}/\mathbb{Z}_{k}$, while the ${\rm O}\times{\rm USp}$ theory comes form the geometry of $\mathbb{C}^{4}/\mathbb{D}_{k}$. The accidental equality $\mathbb{Z}_{4}=\mathbb{D}_{1}$ implies the dualities between the ${\rm U}\left(N_{1}\right)_{4}\times{\rm U}\left(N_{2}\right)_{-4}$ and ${\rm O}\left(N_{1}'\right)_{2}\times{\rm USp}\left(2N_{2}'\right)_{-1}$ theories \cite{Aharony:2008gk}. The concrete correspondence of the ranks of the gauge group is \cite{Cheon:2012be}
\begin{align}
{\rm U}\left(N\right)_{4}\times{\rm U}\left(N\right)_{-4} & \Leftrightarrow{\rm O}\left(2N\right)_{2}\times{\rm USp}\left(2N\right)_{-1},\nonumber \\
{\rm U}\left(N+1\right)_{4}\times{\rm U}\left(N\right)_{-4} & \Leftrightarrow{\rm O}\left(2N+1\right)_{2}\times{\rm USp}\left(2N\right)_{-1},\nonumber \\
{\rm U}\left(N+2\right)_{4}\times{\rm U}\left(N\right)_{-4} & \Leftrightarrow{\rm O}\left(2N+2\right)_{2}\times{\rm USp}\left(2N\right)_{-1}.\label{eq:C4Z4dual}
\end{align}
Since the Seiberg-like duality relates ${\rm U}\left(N+M\right)_{k}\times{\rm U}\left(N\right)_{-k}$ and ${\rm U}\left(N+\left|k\right|-M\right)_{k}\times{\rm U}\left(N\right)_{-k}$,\footnote{Here we also used the parity symmetry $k\leftrightarrow-k$. In terms of the matrix model, this corresponds to the complex conjugate.} these are all the cases of $k=4$. These dualities are important because they come from the geometry of M-theory, and hence they are a strong evidence that the M-theoretical interpretation is correct.

Thanks to the localization technique, now we have a chance to check and study dualities by using the partition functions. Actually, the dualities \eqref{eq:C4Z4dual} were checked by using the superconformal index \cite{Cheon:2012be,Beratto:2021xmn,Hayashi:2022ldo}. However, the check using the superconformal index has been typically restricted to a fixed and small $N$ since the cost for calculating the superconformal index rapidly grows. Furthermore, the agreement is typically checked only for a Taylor series of a fugacity. Finding a decoupled topological sector is also hard in general (an improvement of this point for related theories is discussed in \cite{Beratto:2021xmn}).

On the other hand, the check of the dualities \eqref{eq:C4Z4dual} by using the $S^{3}$ partition function is still missing. In this paper, we study the duality in terms of the $S^{3}$ partition function and improve the situation discussed above. Namely, we prove exact relations between the $S^{3}$ partition functions of the dual theories for arbitrary $N$. These exact relations can be used to check the correspondence of the flavor symmetries between dual theories, which has been found by using the superconformal index \cite{Beratto:2021xmn,Hayashi:2022ldo}, by turning on mass deformations. Moreover, we see that the absolute values of the $S^{3}$ partition functions of the dual theories completely match in a non-trivial way, which implies that no decoupled sector is needed for the dualities.

After applying the supersymmetric localization to the $S^{3}$ partition function, we obtain a matrix model. However, in general, the matrix model is still complicated. For obtaining exact relations for arbitrary $N$, in this paper, we use and develop a computation technique called the Fermi gas formalism. The original idea of the Fermi gas formalism is to rewrite the matrix model as a partition function of an ideal Fermi gas system. The Fermi gas formalism was first introduced in \cite{Marino:2011eh} for $\hat{A}_{n}$-type (necklace) quiver theories. Later, it was generalized to $\hat{D}_{n}$-type quiver theories \cite{Assel:2015hsa,Moriyama:2015jsa} and orientifold theories \cite{Mezei:2013gqa,Moriyama:2015asx,Honda:2015rbb,Okuyama:2016xke}. The Fermi gas formalism for non-uniform rank theories are more difficult than for uniform rank theories in general, and development for this direction was also performed in \cite{Honda:2013pea,Matsumoto:2013nya,Kubo:2020qed} for $\hat{A}_{n}$-type quiver theories and in \cite{Moriyama:2016xin,Moriyama:2016kqi} for the orientifold theories. Roughly speaking, after applying the Fermi gas formalism, almost all information is put into a density matrix of an associated Fermi gas system. Hence, once one can apply the Fermi gas formalism, a remaining task to find exact relations is finding exact relations between the density matrices. In this paper, we actually proceed in this way. Note that in the Fermi gas formalism, the rank $N$ becomes a number of the particles, and hence the result is automatically applicable for arbitrary $N$. This is one advantage of the Fermi gas formalism and the reason why we use this formalism. In other words, the Fermi gas formalism is an efficient tool for studying dualities which are expected to hold for arbitrary $N$ such as \eqref{eq:C4Z4dual}.

We also find an interesting connection between the dualities \eqref{eq:C4Z4dual} and $\hat{A}_{3}=\hat{D}_{3}$. Here $\hat{A}_{n}$ and $\hat{D}_{n}$ are the affine Lie algebras of $\mathfrak{su}\left(n\right)$ and $\mathfrak{so}\left(2n\right)$, respectively, and labeled by the Dynkin diagram. It is known that when $n=3$, they accidentally coincide. The Dynkin diagram parameterizes an interesting class of 3d $\mathcal{N}=4$ quiver theories \cite{Gaiotto:2008sd,Hosomichi:2008jd,Imamura:2008dt}, where the quiver diagram is described by the Dynkin diagram. It was found that the free energy of theories where the shape of quiver diagram is $\hat{A}_{n}$, $\hat{D}_{n}$ or $\hat{E}_{n}$ behaves as $F\sim N^{3/2}$ (when the Chern-Simons levels satisfy the balanced condition, which is always satisfied in this paper) \cite{Herzog:2010hf,Gulotta:2011vp,Gulotta:2012yd,Crichigno:2012sk}. We find that the matrix models for the ${\rm U}\times{\rm U}$ theories (with $k=4$) and the ${\rm O}\times{\rm USp}$ theories (with $k=1$) accidentally coincide with the ones of $\hat{A}_{3}$ and $\hat{D}_{3}$ quiver theories, respectively. This is accidental in the sense that the ${\rm U}\times{\rm U}$ or ${\rm O}\times{\rm USp}$ matrix model corresponds to the $\hat{A}_{3}$ or $\hat{D}_{3}$ matrix model with restricted parameters, respectively, which we will see in section \ref{sec:Dualities}. In other words, the ${\rm U}\times{\rm U}$ or ${\rm O}\times{\rm USp}$ matrix model is embedded in the $\hat{A}_{3}$ or $\hat{D}_{3}$ matrix model, respectively. Since $\hat{A}_{3}=\hat{D}_{3}$, the matrix models for $\hat{A}_{3}$ and $\hat{D}_{3}$ quiver theories are automatically the same. Therefore, by embedding the ${\rm U}\times{\rm U}$ or ${\rm O}\times{\rm USp}$ matrix model, we can obtain the exact relation between the ${\rm U}\times{\rm U}$ matrix model and the ${\rm O}\times{\rm USp}$ matrix model. This is actually the strategy we use in this paper (see figure \ref{fig:Quivers}).

As another application of our result, we will comment on conjectured functional relations between the (normalized) grand partition functions of the ${\rm U}\times{\rm U}$ theories with and without chiral projections which were found in \cite{Grassi:2014uua}. We show that our result is the last piece of a proof of these functional relations.

This paper is organized as follows. In section \ref{sec:Matrix-models}, we introduce the matrix models for the ${\rm U}\times{\rm U}$, $\hat{A}_{3}$, $\hat{D}_{3}$ and ${\rm O}\times{\rm USp}$ quiver theories. In section \ref{sec:FGF}, we apply the Fermi gas formalism to these matrix models. In section \ref{sec:Dualities}, we show exact relations between the matrix models for dual theories. We also comment on the correspondence of the flavor symmetries and decoupled sectors. In section \ref{sec:Functional-Relations}, we comment on the connection between our results and the functional relations found in \cite{Grassi:2014uua}. Finally, in section \ref{sec:Discussion}, we summarize our results and give future directions. In appendix \ref{sec:QM}, we provide notation and useful formulas related to the density matrix appeared in the Fermi gas formalism. In appendix \ref{sec:Formulas}, we list useful formulas for applying the Fermi gas formalism.

\section{Matrix models\label{sec:Matrix-models}}

In this section we introduce the theories and associated matrix models. In this paper, four types of the ABJ(M) theories appear, and they are classified by their gauge group, namely ${\rm U}\times{\rm U}$, $\hat{A}_{3}$, $\hat{D}_{3}$ and ${\rm O}\times{\rm USp}$.
\begin{figure}
\begin{centering}
\includegraphics[scale=0.5]{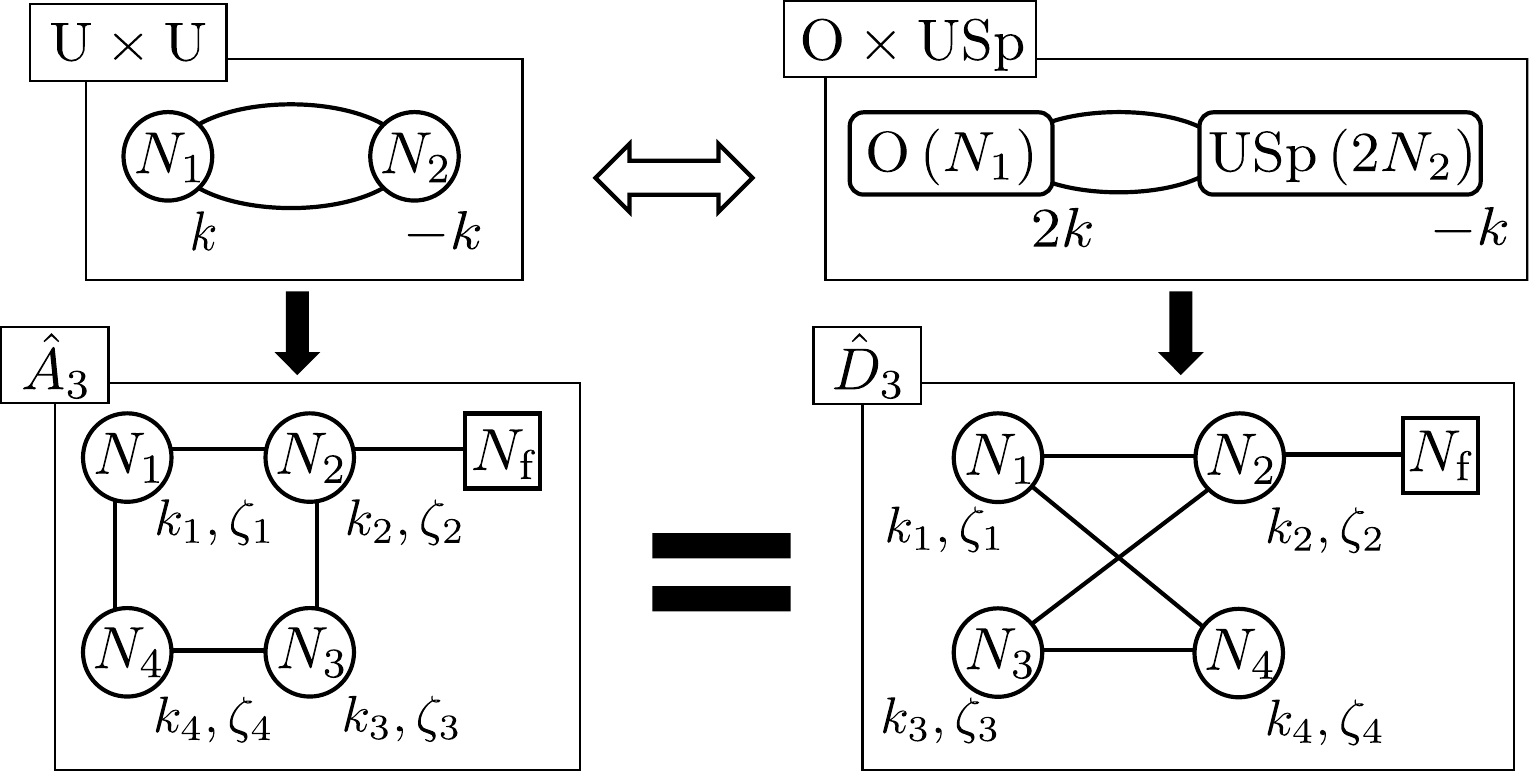}
\par\end{centering}
\caption{$\mathcal{N}=4$ quiver diagrams of 3d Chern-Simons theories which appear in this paper, and schematic relations between them. The upper two theories are dual each other as \eqref{eq:C4Z4dual}. The group of each node for ${\rm U}\times{\rm U}$, $\hat{A}_{3}$ and $\hat{D}_{3}$ are the unitary group. The subscript denotes the Chern-Simons level and the FI parameter. For the ${\rm O}\times{\rm USp}$ quiver, the two of four bi-fundamental fields are projected out by the orientifold projection, though we write two lines. For the $\hat{A}_{3}$ and $\hat{D}_{3}$ quivers, we can also add $N_{{\rm f}}$ fundamental matters.\label{fig:Quivers}}
\end{figure}
The theories can be written in terms of the quiver diagram as in figure \ref{fig:Quivers}. They admit deformations. The ${\rm U}\times{\rm U}$ ABJ(M) theory has ${\rm SU}\left(2\right)\times{\rm SU}\left(2\right)\times{\rm U}\left(1\right)$ flavor symmetry, while the ${\rm O}\times{\rm USp}$ ABJ theory has ${\rm SU}\left(2\right)$ flavor symmetry. It is known from the study of the superconformal index that the diagonal subgroup of the ${\rm SU}\left(2\right)\times{\rm SU}\left(2\right)$ flavor symmetry of ${\rm U}\times{\rm U}$ theory is identified as the ${\rm SU}\left(2\right)$ flavor symmetry of ${\rm O}\times{\rm USp}$ theory \cite{Beratto:2021xmn}. For checking this correspondence in terms of the $S^{3}$ partition function, we turn on these mass deformations for ${\rm U}\times{\rm U}$ and ${\rm O}\times{\rm USp}$ theories. For $\hat{A}_{3}$ and $\hat{D}_{3}$ quivers, we can turn on the FI parameter. We also put $F$ fundamental matters at the second node. In this paper, $F=0$ or $1$. The partition function is parameterized by theses variables as in Figure \ref{fig:Quivers}.

After performing the supersymmetric localization, the partition function on $S^{3}$ becomes a matrix model \cite{Kapustin:2009kz}. The form of the matrix model is
\begin{equation}
Z=\frac{1}{\left|\mathcal{W}\right|}\int d\alpha Z_{\text{cl}}Z_{\text{vec}}Z_{\text{hyp}},
\end{equation}
where $\left|\mathcal{W}\right|$ is the order of the Weyl group. The integration variables come from the Cartan subalgebra of the gauge group. $Z_{\text{cl}}$ is a classical contribution and $Z_{\text{vec}}Z_{\text{hyp}}$ is the 1-loop contribution. In the following, we enumerate these factors more explicitly, where the integration variables are rescaled as $\alpha\rightarrow\alpha/2\pi$. The classical contribution comes from the Chern-Simons terms and the FI terms
\begin{equation}
Z_{\text{cl}}\rightarrow e^{\frac{ik}{4\pi}{\rm Tr}\alpha^{2}+i\zeta{\rm Tr}\alpha},
\end{equation}
where $k$ and $\zeta$ are the Chern-Simons level and the FI parameter of each node, respectively. ${\rm Tr}$ is the trace in the fundamental representation. The contribution to the 1-loop part comes from the vector multiplet and the hypermultiplet
\begin{equation}
Z_{\text{vec}}\rightarrow\prod_{\gamma>0}\left(2\sinh\frac{\gamma\cdot\alpha}{2}\right)^{2},\quad Z_{\text{hyp}}\rightarrow\prod_{w\in{\cal R}}\frac{1}{2\cosh\frac{w\cdot\alpha+2\pi m}{2}},
\end{equation}
where $\gamma$ runs over the positive roots of the Lie algebra and $w$ runs over the weights of ${\cal R}$. ${\cal R}$ is the representation of the hypermultiplet.
\begin{table}
\begin{centering}
\begin{tabular}{|c|c|}
\hline 
Vector multiplet & 1-loop factor\tabularnewline
\hline 
\hline 
${\rm U}\left(N\right)$ & $\prod_{a<a'}^{N}\left(2\sinh\frac{\alpha_{a}-\alpha_{a'}}{2}\right)^{2}$\tabularnewline
\hline 
${\rm O}\left(2N\right)$ & $\prod_{a<a'}^{N}\left(2\sinh\frac{\alpha_{a}-\alpha_{a'}}{2}2\sinh\frac{\alpha_{a}+\alpha_{a'}}{2}\right)^{2}$\tabularnewline
\hline 
${\rm O}\left(2N+1\right)$ & $\prod_{a}^{N}\left(2\sinh\frac{\alpha_{a}}{2}\right)^{2}\prod_{a<a'}^{N}\left(2\sinh\frac{\alpha_{a}-\alpha_{a'}}{2}2\sinh\frac{\alpha_{a}+\alpha_{a'}}{2}\right)^{2}$\tabularnewline
\hline 
${\rm USp}\left(2N\right)$ & $\prod_{a}^{N}\left(2\sinh\alpha_{a}\right)^{2}\prod_{a<a'}^{N}\left(2\sinh\frac{\alpha_{a}-\alpha_{a'}}{2}2\sinh\frac{\alpha_{a}+\alpha_{a'}}{2}\right)^{2}$\tabularnewline
\hline 
\end{tabular}
\par\end{centering}
\medskip{}

\begin{centering}
\begin{tabular}{|c|c|}
\hline 
{\footnotesize{}(Bi-)fundamental hypermultiplet} & 1-loop factor\tabularnewline
\hline 
\hline 
${\rm U}\left(N\right)$ (massless) & $\frac{1}{\prod_{a}^{N}2\cosh\frac{\alpha_{a}}{2}}$\tabularnewline
\hline 
${\rm U}\left(N_{1}\right)\times{\rm U}\left(N_{2}\right)$ & $\frac{1}{\prod_{a,b}^{N_{1},N_{2}}2\cosh\frac{\alpha_{a}-\beta_{b}+2\pi m}{2}}$\tabularnewline
\hline 
${\rm O}\left(2N_{1}\right)\times{\rm USp}\left(2N_{2}\right)$ & $\frac{1}{\prod_{a,b}^{N_{1},N_{2}}\prod_{\epsilon=\pm}2\cosh\frac{\alpha_{a}-\beta_{b}+2\pi m\epsilon}{2}2\cosh\frac{\alpha_{a}+\beta_{b}+2\pi m\epsilon}{2}}$\tabularnewline
\hline 
${\rm O}\left(2N_{1}+1\right)\times{\rm USp}\left(2N_{2}\right)$ & $\frac{1}{\prod_{a,b}^{N_{1},N_{2}}\prod_{\epsilon=\pm}2\cosh\frac{\alpha_{a}-\beta_{b}+2\pi m\epsilon}{2}2\cosh\frac{\alpha_{a}+\beta_{b}+2\pi m\epsilon}{2}}\frac{1}{\prod_{b}^{N_{2}}\prod_{\epsilon=\pm}2\cosh\frac{\beta_{b}+2\pi m\epsilon}{2}}$\tabularnewline
\hline 
\end{tabular}
\par\end{centering}
\caption{The 1-loop contributions from the vector multiplet and the (bi-)fundamental hypermultiplet. The factor coming from ${\rm O}\times{\rm USp}$ is the whole effect of two hypermultiplet with orientifold projection.\label{tab:Vector}}
\end{table}
Table \ref{tab:Vector} shows the contribution from the vector multiplet and the hypermultiplet \cite{Gulotta:2012yd,Assel:2015hsa}. Note that in this paper we use the following notation for productions
\begin{equation}
\prod_{a}^{N}=\prod_{a=1}^{N},\quad\prod_{a,b}^{N_{1},N_{2}}=\prod_{a=1}^{N_{1}}\prod_{b=1}^{N_{2}},\quad\prod_{a<b}^{N}=\prod_{a=1}^{N-1}\prod_{b=a+1}^{N}.
\end{equation}

Now we explicitly provide the matrix models for ${\rm U}\times{\rm U}$, $\hat{A}_{3}$, $\hat{D}_{3}$ and ${\rm O}\times{\rm USp}$ theories with the parameters shown in figure \ref{fig:Quivers}. The matrix model for ${\rm U}\left(N_{1}\right)_{k}\times{\rm U}\left(N_{2}\right)_{-k}$ is\footnote{In this paper, the range of integration is always from $-\infty$ to $+\infty$.}
\begin{align}
Z_{k,\left(N_{1},N_{2}\right)}^{{\rm U}\times{\rm U}}\left(m\right) & =\frac{i^{-\frac{1}{2}\left(N_{1}^{2}-N_{2}^{2}\right)}}{N_{1}!N_{2}!}\int\frac{d^{N_{1}}\mu}{\left(2\pi\right)^{N_{1}}}\frac{d^{N_{2}}\nu}{\left(2\pi\right)^{N_{2}}}e^{\frac{ik}{4\pi}\left(\sum_{a}^{N_{1}}\mu_{a}^{2}-\sum_{b}^{N_{2}}\nu_{b}^{2}\right)}\nonumber \\
 & \quad\times\frac{\prod_{a<a'}^{N_{1}}\left(2\sinh\frac{\mu_{a}-\mu_{a'}}{2}\right)^{2}\prod_{b<b'}^{N_{2}}\left(2\sinh\frac{\nu_{b}-\nu_{b'}}{2}\right)^{2}}{\prod_{a,b}^{N_{1},N_{2}}2\cosh\frac{\mu_{a}-\nu_{b}+2\pi m}{2}\prod_{b,a}^{N_{2}N_{1}}2\cosh\frac{\nu_{b}-\mu_{a}+2\pi m}{2}}.\label{eq:PF_UU}
\end{align}
The mass parameter $m$ corresponds to the diagonal subgroup of the ${\rm SU}\left(2\right)\times{\rm SU}\left(2\right)$ flavor symmetry as commented above. The matrix model for $\hat{A}_{3}$ quiver diagram with the parameters
\begin{equation}
\bm{k}=\left(k_{1},k_{2},k_{3},k_{4}\right),\quad\bm{N}=\left(N_{1},N_{2},N_{3},N_{4}\right),\quad\bm{\zeta}=\left(\zeta_{1},\zeta_{2},\zeta_{3},\zeta_{4}\right),
\end{equation}
is
\begin{align}
Z_{\bm{k},\bm{N},F}^{\hat{A}_{3}}\left(\bm{\zeta}\right) & =\prod_{r=1}^{4}\frac{i^{-\frac{1}{2}{\rm sgn}\left(k_{r}\right)N_{r}^{2}}}{N_{r}!}\int\prod_{r=1}^{4}\frac{d^{N_{r}}\alpha^{\left(r\right)}}{\left(2\pi\right)^{N_{r}}}\prod_{r=1}^{4}e^{\frac{ik_{r}}{4\pi}\sum_{a}^{N_{r}}\left(\alpha_{a}^{\left(r\right)}\right)^{2}+i\zeta_{r}\sum_{a}^{N_{r}}\alpha_{a}^{\left(r\right)}}\nonumber \\
 & \quad\times\prod_{r=1}^{4}\frac{\prod_{a<a'}^{N_{r}}\left(2\sinh\frac{\alpha_{a}^{\left(r\right)}-\alpha_{a'}^{\left(r\right)}}{2k}\right)^{2}}{\prod_{a,b}^{N_{r},N_{r+1}}2\cosh\frac{\alpha_{a}^{\left(r\right)}-\alpha_{b}^{\left(r+1\right)}}{2}}\left(\frac{1}{\prod_{a}^{N_{2}}2\cosh\frac{\alpha_{a}^{\left(2\right)}}{2}}\right)^{F},\label{eq:PF_A3}
\end{align}
where $\alpha^{\left(5\right)}=\alpha^{\left(1\right)}$, $N_{5}=N_{1}$ and
\begin{equation}
{\rm sgn}\left(k\right)=\begin{cases}
1 & \left(k>0\right)\\
0 & \left(k=0\right)\\
-1 & \left(k<0\right)
\end{cases}.
\end{equation}
The matrix model for the $\hat{D}_{3}$ quiver theory with the same parameters is the same with the one for $\hat{A}_{3}$ quiver theory
\begin{equation}
Z_{\bm{k},\bm{N},F}^{\hat{D}_{3}}\left(\bm{\zeta}\right)=Z_{\bm{k},\bm{N},F}^{\hat{A}_{3}}\left(\bm{\zeta}\right).\label{eq:PF_D3}
\end{equation}
The matrix model for ${\rm O}\left(2N_{1}\right)_{2k}\times{\rm USp}\left(2N_{2}\right)_{-k}$ is
\begin{align}
 & Z_{k,\left(2N_{1},2N_{2}\right)}^{{\rm O}\times{\rm USp}}\left(m\right)\nonumber \\
 & =\frac{i^{-\left(N_{1}^{2}-\frac{1}{2}N_{1}\right)+\left(N_{2}^{2}+\frac{1}{2}N_{2}\right)}}{2^{N_{1}+N_{2}}N_{1}!N_{2}!}\int\frac{d^{N_{1}}\mu}{\left(2\pi\right)^{N_{1}}}\frac{d^{N_{2}}\nu}{\left(2\pi\right)^{N_{2}}}e^{\frac{ik}{2\pi}\left(\sum_{a}^{N_{1}}\mu_{a}^{2}-\sum_{b}^{N_{2}}\nu_{b}^{2}\right)}\prod_{b}^{N_{2}}\left(2\sinh\nu_{b}\right)^{2}\nonumber \\
 & \quad\times\frac{\prod_{a<a'}^{N_{1}}\left(2\sinh\frac{\mu_{a}-\mu_{a'}}{2}2\sinh\frac{\mu_{a}+\mu_{a'}}{2}\right)^{2}\prod_{b<b'}^{N_{2}}\left(2\sinh\frac{\nu_{b}-\nu_{b'}}{2}2\sinh\frac{\nu_{b}+\nu_{b'}}{2}\right)^{2}}{\prod_{a,b}^{N_{1},N_{2}}\left(2\cosh\frac{\mu_{a}-\nu_{b}+2\pi m}{2}\right)\left(2\cosh\frac{\mu_{a}+\nu_{b}+2\pi m}{2}\right)\left(2\cosh\frac{\mu_{a}-\nu_{b}-2\pi m}{2}\right)\left(2\cosh\frac{\mu_{a}+\nu_{b}-2\pi m}{2}\right)}.\label{eq:PF_OUSpE}
\end{align}
The matrix model for ${\rm O}\left(2N_{1}+1\right)_{2k}\times{\rm USp}\left(2N_{2}\right)_{-k}$ is
\begin{align}
 & Z_{k,\left(2N_{1}+1,2N_{2}\right)}^{{\rm O}\times{\rm USp}}\left(m\right)\nonumber \\
 & =\frac{i^{-\left(N_{1}^{2}+\frac{1}{2}N_{1}\right)+\left(N_{2}^{2}+\frac{1}{2}N_{2}\right)}}{2^{N_{1}+N_{2}+1}N_{1}!N_{2}!}\int\frac{d^{N_{1}}\mu}{\left(2\pi\right)^{N_{1}}}\frac{d^{N_{2}}\nu}{\left(2\pi\right)^{N_{2}}}e^{\frac{ik}{2\pi}\left(\sum_{a}^{N_{1}}\mu_{a}^{2}-\sum_{b}^{N_{2}}\nu_{b}^{2}\right)}\frac{\prod_{a}^{N_{1}}\left(2\sinh\frac{\mu_{a}}{2}\right)^{2}\prod_{b}^{N_{2}}\left(2\sinh\nu_{b}\right)^{2}}{\prod_{b}^{N_{2}}\left(2\cosh\frac{\nu_{b}+2\pi m}{2}\right)\left(2\cosh\frac{\nu_{b}-2\pi m}{2}\right)}\nonumber \\
 & \quad\times\frac{\prod_{a<a'}^{N_{1}}\left(2\sinh\frac{\mu_{a}-\mu_{a'}}{2}2\sinh\frac{\mu_{a}+\mu_{a'}}{2}\right)^{2}\prod_{b<b'}^{N_{2}}\left(2\sinh\frac{\nu_{b}-\nu_{b'}}{2}2\sinh\frac{\nu_{b}+\nu_{b'}}{2}\right)^{2}}{\prod_{a,b}^{N_{1},N_{2}}\left(2\cosh\frac{\mu_{a}-\nu_{b}+2\pi m}{2}\right)\left(2\cosh\frac{\mu_{a}+\nu_{b}+2\pi m}{2}\right)\left(2\cosh\frac{\mu_{a}-\nu_{b}-2\pi m}{2}\right)\left(2\cosh\frac{\mu_{a}+\nu_{b}-2\pi m}{2}\right)}.\label{eq:PF_OUSpO}
\end{align}
For both the ${\rm O}\times{\rm USp}$ cases, the mass parameter $m$ corresponds to the ${\rm SU}\left(2\right)$ flavor symmetry as commented above. Note that the orders of the Weyl groups of ${\rm SO}\left(2L\right)$ and ${\rm SO}\left(2L\right)$ are $2^{L-1}L!$ and $2^{L}L!$, respectively. However, since the group is ${\rm O}$, we need an additional factor $\frac{1}{2}$ coming from the gauging of the $\mathbb{Z}_{2}$ charge conjugation symmetry \cite{Mezei:2013gqa}. Note also that the phase factors for ${\rm U}\times{\rm U}$, $\hat{A}_{3}$, $\hat{D}_{3}$ and ${\rm O}\times{\rm USp}$ matrix models are chosen so that when $N_{1}=0$ or $N_{2}=0$ the matrix model reduces to the one of the pure Chern-Simons theory with corresponding gauge group (see \cite{Marino:2011nm}, for example).

\section{Fermi gas formalism\label{sec:FGF}}

In this section we apply the Fermi gas formalism to the matrix models for the four types of the quiver theories: ${\rm U}\times{\rm U}$, $\hat{A}_{3}$, $\hat{D}_{3}$ and ${\rm O}\times{\rm USp}$. The detailed computation sometimes strongly depends on the parameter of the theory, especially the Chern-Simons levels and the ranks, hence we will study different parameters separately when necessary.

As we will see in this section, after applying the Fermi gas formalism, almost all information of the matrix model is put into the density matrix $\hat{\rho}_{k}\left(\hat{x},\hat{p}\right)$ of an associated ideal Fermi gas system (the density matrix is labeled by additional parameters in general). In this paper, the commutation relation between $\hat{x}$ and $\hat{p}$ always satisfies
\begin{equation}
\left[\hat{x},\hat{p}\right]=\begin{cases}
2\pi ik & \left(k>0\right)\\
2\pi i & \left(k=0\right)
\end{cases}.
\end{equation}

After obtaining the partition function of an associated ideal Fermi gas system, we also introduce the grand partition function which is useful and sometimes crucial to find the relation between different matrix models. To move from the canonical ensemble to the grand canonical ensemble, we interpret the overall rank as a number of particles and introduce fugacity $\kappa$. 

\subsection{${\rm U}\left(N+M\right)_{k}\times{\rm U}\left(N\right)_{-k}$ theory\label{subsec:FGF-UU}}

In this section we review the Fermi gas formalism for the matrix model for the mass deformed ABJ theory. This was performed in \cite{Nosaka:2015iiw}. One motivation of this section is that, because the strategy of computation here can be applicable also for the other matrix models, this section would be useful for following the Fermi gas formalism in this paper. Here we reparametrize as $\left(N_{1},N_{2}\right)=\left(N+M,N\right)$ and we assume $M>0$. Roughly speaking, for applying the Fermi gas formalism we need to perform the integration over the deformed rank (in this case $M$). We achieve this by generating $M$ delta functions.

The matrix model is defined in \eqref{eq:PF_UU}. We rescale the integration variables as $\alpha\rightarrow\alpha/k$. We first apply the determinant formula \eqref{eq:DetFormula-cosh} to the 1-loop determinant factors with setting $\hbar=2\pi k$. Note that for applying the determinant formula, we need to shift $\mu$ or $\nu$ in \eqref{eq:DetFormula-cosh} by the mass parameter appropriately. We also put the Chern-Simons factor to the first determinant as
\begin{align}
 & Z_{k,\left(N+M,N\right)}^{{\rm U}\times{\rm U}}\left(m\right)\nonumber \\
 & =\frac{i^{-MN-\frac{1}{2}M^{2}}}{\hbar^{2N+M}}\int\frac{d^{N+M}\mu}{\left(N+M\right)!}\frac{d^{N}\nu}{N!}\det\left(\begin{array}{c}
\left[\hbar\braket{\nu_{a}|e^{-\frac{i}{4\pi k}\hat{x}^{2}}\frac{e^{im\hat{p}}}{2\cosh\frac{\hat{p}-i\pi M}{2}}e^{\frac{i}{4\pi k}\hat{x}^{2}}|\mu_{b}}\right]_{a,b}^{N\times\left(N+M\right)}\\
\left[\frac{\hbar}{\sqrt{k}}\bbraket{t_{M,j}|e^{\frac{i}{4\pi k}\hat{x}^{2}}|\mu_{b}}\right]_{j,b}^{M\times\left(N+M\right)}
\end{array}\right)\nonumber \\
 & \quad\times\det\left(\begin{array}{cc}
\left[\hbar\braket{\mu_{a}|\frac{e^{im\hat{p}}}{2\cosh\frac{\hat{p}+i\pi M}{2}}|\nu_{b}}\right]_{a,b}^{\left(N+M\right)\times N} & \left[\frac{\hbar}{\sqrt{k}}\brakket{\mu_{a}|-t_{M,j}}\right]_{a,j}^{\left(N+M\right)\times M}\end{array}\right).
\end{align}
The first determinant can be diagonalized since the following identity holds for an anti-symmetric function $g\left(\vec{\alpha}\right)$
\begin{equation}
\frac{1}{N!}\int d^{N}\alpha\det\left(\left[f_{a}\left(\alpha_{b}\right)\right]_{a,b}^{N\times N}\right)g\left(\alpha_{1},\ldots,\alpha_{N}\right)=\int d^{N}\alpha\left(\prod_{a=1}^{N}f_{a}\left(\alpha_{a}\right)\right)g\left(\alpha_{1},\ldots,\alpha_{N}\right).\label{eq:DiagFormula}
\end{equation}
Now we can perform the similarity transformations
\begin{equation}
\int d\mu\ket{\mu}\bra{\mu}=\int d\mu e^{\frac{i}{4\pi k}\hat{p}^{2}}\ket{\mu}\bra{\mu}e^{-\frac{i}{4\pi k}\hat{p}^{2}},\quad\int d\mu\ket{\nu}\bra{\nu}=\int d\nu e^{\frac{i}{4\pi k}\hat{p}^{2}}\ket{\nu}\bra{\nu}e^{-\frac{i}{4\pi k}\hat{p}^{2}}.
\end{equation}
Thanks to the formula \eqref{eq:OpSim}, the delta function $\delta\left(\mu_{N+j}-t_{M,j}\right)$ ($j=1,2,\ldots,M$) appears at the lower row of the first determinant. An important point is that because $\hat{x}$ operator does not appear in the second determinant, the similarity transformation does not affect to the second determinant (up to a phase factor coming from the momentum vector). Hence we can apply the determinant formula \eqref{eq:DetFormula-cosh} backwards to the second determinant. We then perform the integration over $\mu_{N+j}$ ($j=1,2,\ldots,M$) by using the delta functions. Now we arrive at
\begin{align}
Z_{k,\left(N+M,N\right)}^{{\rm U}\times{\rm U}}\left(m\right) & =\frac{i^{-MN}}{\hbar^{N}}Z_{k,M}^{{\rm U}}\left(0\right)\int d^{N}\mu\frac{d^{N}\nu}{N!}\prod_{a=1}^{N}\left(\braket{\nu_{a}|\frac{e^{im\hat{x}}}{2\cosh\frac{\hat{x}-i\pi M}{2}}|\mu_{a}}\prod_{j}^{M}2\sinh\frac{\mu_{a}-t_{M,j}}{2k}\right)\nonumber \\
 & \quad\times\frac{\prod_{a<a'}^{N}2\sinh\frac{\mu_{a}-\mu_{a'}}{2k}\prod_{b<b'}^{N}2\sinh\frac{\nu_{b}-\nu_{b'}}{2k}}{\prod_{a,b}^{N,N}2\cosh\frac{\mu_{a}-\nu_{b}+2\pi km}{2k}}\frac{1}{\prod_{j}^{M}\prod_{b}^{N}2\cosh\frac{t_{M,j}-\nu_{b}+2\pi km}{2k}},
\end{align}
where $Z_{k,M}^{{\rm U}}\left(\zeta\right)$ is the $S^{3}$ partition function of the pure Chern-Simons theory with ${\rm U}\left(M\right)_{k}$ gauge group defined by
\begin{equation}
Z_{k,M}^{{\rm U}}\left(\zeta\right)=e^{\frac{\pi i}{6k}\left(M^{3}-M\right)-\frac{\pi i}{k}M\zeta^{2}}\frac{1}{k^{\frac{M}{2}}}\prod_{j<j'}^{M}2\sin\frac{\pi}{k}\left(j'-j\right).\label{eq:Z_U-Def}
\end{equation}
Here, for later convenience, we introduced an extra parameter $\zeta$, which will be related to a FI parameter. We again apply the determinant formula \eqref{eq:DetFormula-cosh} and put the remaining integrand to the determinant as
\begin{align}
 & Z_{k,\left(N+M,N\right)}^{{\rm U}\times{\rm U}}\left(m\right)\nonumber \\
 & =i^{-MN}Z_{k,M}^{{\rm U}}\left(0\right)\int d^{N}\mu\frac{d^{N}\nu}{N!}\nonumber \\
 & \quad\times\det\left(\left[\braket{\nu_{a}|e^{im\hat{x}}\frac{\prod_{j}^{M}2\sinh\frac{\hat{x}-t_{M,j}}{2k}}{2\cosh\frac{\hat{x}-i\pi M}{2}}|\mu_{a}}\braket{\mu_{a}|\frac{e^{im\hat{p}}}{2\cosh\frac{\hat{p}}{2}}\frac{1}{\prod_{j}^{M}2\cosh\frac{\hat{x}-2\pi km-t_{M,j}}{2k}}|\nu_{b}}\right]_{a,b}^{N\times N}\right).
\end{align}
The integration over $\mu$ is now trivial since $\int d\mu\ket{\mu}\bra{\mu}=1$. Hence, after putting $i^{-MN}$ to the determinant and relabeling $\nu\rightarrow\mu$, we finally arrive at
\begin{equation}
Z_{k,\left(N+M,N\right)}^{{\rm U}\times{\rm U}}\left(m\right)=Z_{k,M}^{{\rm U}}\left(0\right)\int\frac{d^{N}\mu}{N!}\det\left(\left[\braket{\mu_{a}|\hat{\rho}_{k,M}^{{\rm U}\times{\rm U}}\left(\hat{x},\hat{p};m\right)|\mu_{b}}\right]_{a,b}^{N\times N}\right),\label{eq:PF-UU-FGF}
\end{equation}
where the density matrix is defined by
\begin{equation}
\hat{\rho}_{k,M}^{{\rm U}\times{\rm U}}\left(\hat{x},\hat{p};m\right)=e^{im\hat{x}}\frac{\prod_{j}^{M}2\sinh\frac{\hat{x}-t_{M,j}}{2k}}{e^{\frac{\hat{x}}{2}}+\left(-1\right)^{M}e^{-\frac{\hat{x}}{2}}}\frac{1}{2\cosh\frac{\hat{p}}{2}}\frac{1}{\prod_{j}^{M}2\cosh\frac{\hat{x}-t_{M,j}}{2k}}e^{im\hat{p}}.\label{eq:DM-UU}
\end{equation}

Finally, we introduce the grand partition function which is defined as
\begin{equation}
\Xi_{k,M}^{{\rm U}\times{\rm U}}\left(\kappa;m\right)=\sum_{N=0}^{\infty}\kappa^{N}Z_{k,\left(N+M,N\right)}^{{\rm U}\times{\rm U}}\left(m\right).\label{eq:GPF-UU-def}
\end{equation}
According to the formula \eqref{eq:GPF-Circle}, we obtain
\begin{equation}
\Xi_{k,M}^{{\rm U}\times{\rm U}}\left(\kappa;m\right)=Z_{k,M}^{{\rm U}}\left(0\right){\rm Det}\left(1+\kappa\hat{\rho}_{k,M}^{{\rm U}\times{\rm U}}\left(\hat{x},\hat{p};m\right)\right).\label{eq:GPF-UU-FGF}
\end{equation}

\subsection{$\hat{A}_{3}$ theory\label{subsec:FGF-A3}}

In this section we apply the Fermi gas formalism to the matrix model for the $\hat{A}_{3}$ theory defined in \eqref{eq:PF_A3}. The flow of the computation is similar to the ${\rm U}\times{\rm U}$ case in section \ref{subsec:FGF-UU}.

We consider the case of
\begin{equation}
\left(k_{1},k_{2},k_{3},k_{4}\right)=\left(-k,k,-k,k\right),\quad\left(N_{1},N_{2},N_{3},N_{4}\right)=\left(N,N,N,N+M\right),\label{eq:A3-para}
\end{equation}
with $M\geq0$. For this parameterization, we introduce the grand partition function which is defined as
\begin{equation}
\Xi_{k,M,F}^{\hat{A}_{3}}\left(\kappa;\bm{\zeta}\right)=\sum_{N=0}^{\infty}\kappa^{N}Z_{\bm{k},\bm{N},F}^{\hat{A}_{3}}\left(\bm{\zeta}\right).\label{eq:GPF-A3-def}
\end{equation}
In the following we further restrict our attention to two cases.

\subsubsection{$F=0$ case}

The first case is when $F=0$ and $k>0$. After rescaling the integration variables to $\alpha\rightarrow\alpha/k$, we apply the determinant formula \eqref{eq:DetFormula-cosh} to the 1-loop determinant factors with setting $\hbar=2\pi k$. We also put the Chern-Simons factor and the FI factor to the determinants appropriately as
\begin{align}
Z_{\bm{k},\bm{N},F}^{\hat{A}_{3}}\left(\bm{\zeta}\right) & =\frac{i^{-MN-\frac{1}{2}M^{2}}}{\hbar^{4N+M}}\int\frac{d^{N}\mu}{N!}\frac{d^{N}\nu}{N!}\frac{d^{N}\rho}{N!}\frac{d^{N+M}\sigma}{\left(N+M\right)!}\nonumber \\
 & \quad\times\det\left(\left[\hbar\braket{\mu_{a}|e^{\frac{i\zeta_{1}}{k}\hat{x}}e^{-\frac{i}{4\pi k}\hat{x}^{2}}\frac{1}{2\cosh\frac{\hat{p}}{2}}e^{\frac{i}{4\pi k}\hat{x}^{2}}e^{\frac{i\zeta_{2}}{k}\hat{x}}|\nu_{b}}\right]_{a,b}^{N\times N}\right)\nonumber \\
 & \quad\times\det\left(\left[\hbar\braket{\nu_{a}|\frac{1}{2\cosh\frac{\hat{p}}{2}}|\rho_{b}}\right]_{a,b}^{N\times N}\right)\nonumber \\
 & \quad\times\det\left(\begin{array}{c}
\left[\hbar\braket{\rho_{a}|e^{\frac{i\zeta_{3}}{k}\hat{x}}e^{-\frac{i}{4\pi k}\hat{x}^{2}}\frac{1}{2\cosh\frac{\hat{p}-i\pi M}{2}}e^{\frac{i}{4\pi k}\hat{x}^{2}}e^{\frac{i\zeta_{4}}{k}\hat{x}}|\sigma_{b}}\right]_{a,b}^{N\times\left(N+M\right)}\\
\left[\frac{\hbar}{\sqrt{k}}\bbraket{t_{M,j}|e^{\frac{i}{4\pi k}\hat{x}^{2}}e^{\frac{i\zeta_{4}}{k}\hat{x}}|\sigma_{b}}\right]_{j,b}^{M\times\left(N+M\right)}
\end{array}\right)\nonumber \\
 & \quad\times\det\left(\begin{array}{cc}
\left[\hbar\braket{\sigma_{a}|\frac{1}{2\cosh\frac{\hat{p}+i\pi M}{2}}|\mu_{b}}\right]_{a,b}^{\left(N+M\right)\times N} & \left[\frac{\hbar}{\sqrt{k}}\brakket{\sigma_{a}|-t_{M,j}}\right]_{a,j}^{\left(N+M\right)\times M}\end{array}\right).\label{eq:PF-A3-c1}
\end{align}
The third determinant can be diagonalized by using the formula \eqref{eq:DiagFormula}. In this expression, we can perform the similarity transformations
\begin{equation}
\int d\alpha\ket{\alpha}\bra{\alpha}=\int d\alpha e^{\frac{i}{2\hbar}\hat{p}^{2}}\ket{\alpha}\bra{\alpha}e^{-\frac{i}{2\hbar}\hat{p}^{2}},
\end{equation}
for $\alpha=\mu,\nu,\rho,\sigma$. Thanks to the formula \eqref{eq:OpSim}, the delta functions $\delta\left(\sigma_{N+j}+2\pi\zeta_{4}-t_{M,j}\right)$ ($j=1,2,\ldots,M$) appear at the lower row of the third determinant. After applying the determinant formula \eqref{eq:DetFormula-cosh} backwards to the fourth determinant, we can perform the integration over $\sigma_{N+j}$ ($j=1,2,\ldots,M$) by using the delta functions. We then apply the determinant formula \eqref{eq:DetFormula-cosh} again to the factor coming from the fourth determinant and put the remaining factors to the determinant appropriately. The integration over $\rho$ is now trivial since $\int d\sigma\ket{\sigma}\bra{\sigma}=1$. Therefore, after putting $i^{-MN}$ to the determinant, we obtain
\begin{align}
 & Z_{\bm{k},\bm{N},0}^{\hat{A}_{3}}\left(\bm{\zeta}\right)\nonumber \\
 & =Z_{k,M}^{{\rm U}}\left(\zeta_{4}\right)\int\frac{d^{N}\mu}{N!}\frac{d^{N}\nu}{N!}\frac{d^{N}\rho}{N!}\nonumber \\
 & \quad\times\det\left(\left[\braket{\mu_{a}|e^{\frac{i\zeta_{1}}{k}\left(\hat{x}-\hat{p}\right)}\frac{1}{2\cosh\frac{\hat{x}}{2}}e^{\frac{i\zeta_{2}}{k}\left(\hat{x}-\hat{p}\right)}|\nu_{b}}\right]_{a,b}^{N\times N}\right)\det\left(\left[\braket{\nu_{a}|\frac{1}{2\cosh\frac{\hat{p}}{2}}|\rho_{b}}\right]_{a,b}^{N\times N}\right)\nonumber \\
 & \quad\times\det\left(\left[\braket{\rho_{a}|e^{\frac{i\zeta_{3}}{k}\left(\hat{x}-\hat{p}\right)}\frac{\prod_{j}^{M}2\sinh\frac{\hat{x}-t_{M,j}}{2k}}{e^{\frac{\hat{x}}{2}}+\left(-1\right)^{M}e^{-\frac{\hat{x}}{2}}}e^{\frac{i\zeta_{4}}{k}\left(\hat{x}-\hat{p}\right)}\frac{1}{2\cosh\frac{\hat{p}}{2}}\frac{1}{\prod_{j}^{M}2\cosh\frac{\hat{x}+2\pi\zeta_{4}-t_{M,j}}{2k}}|\mu_{b}}\right]_{a,b}^{N\times N}\right),
\end{align}
where $Z_{k,M}^{{\rm U}}$ is defined in \eqref{eq:Z_U-Def}. By using the gluing formula
\begin{equation}
\int\frac{d^{N}\beta}{N!}\det\left(\left[\braket{\alpha_{a}|\hat{{\cal O}}_{1}|\beta_{b}}\right]_{a,b}^{N\times N}\right)\det\left(\left[\braket{\beta_{a}|\hat{{\cal O}}_{2}|\gamma_{b}}\right]_{a,b}^{N\times N}\right)=\det\left(\left[\braket{\alpha_{a}|\hat{{\cal O}}_{1}\hat{{\cal O}}_{2}|\gamma_{b}}\right]_{a,b}^{N\times N}\right),\label{eq:Glue-Formula}
\end{equation}
we finally arrive at
\begin{equation}
Z_{\bm{k},\bm{N},0}^{\hat{A}_{3}}\left(\bm{\zeta}\right)=Z_{k,M}^{{\rm U}}\left(\zeta_{4}\right)\int\frac{d^{N}\mu}{N!}\det\left(\left[\braket{\mu_{a}|\hat{\rho}_{k,M,0}^{\hat{A}_{3}}\left(\hat{x},\hat{p};\bm{\zeta}\right)|\mu_{b}}\right]_{a,b}^{N\times N}\right),\label{eq:PF-A3-FGF1}
\end{equation}
where the density matrix is defined by
\begin{align}
\hat{\rho}_{k,M,0}^{\hat{A}_{3}}\left(\hat{x},\hat{p};\bm{\zeta}\right) & =e^{\frac{i\zeta_{1}}{k}\left(\hat{x}-\hat{p}\right)}\frac{1}{2\cosh\frac{\hat{x}}{2}}e^{\frac{i\zeta_{2}}{k}\left(\hat{x}-\hat{p}\right)}\frac{1}{2\cosh\frac{\hat{p}}{2}}\nonumber \\
 & \quad\times e^{\frac{i\zeta_{3}}{k}\left(\hat{x}-\hat{p}\right)}\frac{\prod_{j}^{M}2\sinh\frac{\hat{x}-t_{M,j}}{2k}}{e^{\frac{\hat{x}}{2}}+\left(-1\right)^{M}e^{-\frac{\hat{x}}{2}}}e^{\frac{i\zeta_{4}}{k}\left(\hat{x}-\hat{p}\right)}\frac{1}{2\cosh\frac{\hat{p}}{2}}\frac{1}{\prod_{j}^{M}2\cosh\frac{\hat{x}+2\pi\zeta_{4}-t_{M,j}}{2k}}.\label{eq:DM-A31}
\end{align}
Here $\bm{k}$ and $\bm{N}$ are defined in \eqref{eq:A3-para}.

Finally, we consider the grand partition function defined in \eqref{eq:GPF-A3-def}. According to the formula \eqref{eq:GPF-Circle}, we obtain
\begin{equation}
\Xi_{k,M,0}^{\hat{A}_{3}}\left(\kappa;\bm{\zeta}\right)=Z_{k,M}^{{\rm U}}\left(\zeta_{4}\right){\rm Det}\left(1+\kappa\hat{\rho}_{k,M,0}^{\hat{A}_{3}}\left(\hat{x},\hat{p};\bm{\zeta}\right)\right).\label{eq:GPF-A3-FGF1}
\end{equation}

\subsubsection{$k=0$ and $M=0$ case}

The second case is when $k=0$ and $M=0$. We apply the determinant formula \eqref{eq:DetFormula-cosh} to the 1-loop determinant factors with setting $\hbar=2\pi$. We also put the FI factors to the determinants appropriately. We also put the 1-loop determinant factor of the fundamental matter to the first determinant.
\begin{align}
Z_{\bm{k},\bm{N},F}^{\hat{A}_{3}}\left(\bm{\zeta}\right) & =\frac{1}{\hbar^{4N}}\int\frac{d^{N}\mu}{N!}\frac{d^{N}\nu}{N!}\frac{d^{N}\rho}{N!}\frac{d^{N}\sigma}{N!}\nonumber \\
 & \quad\times\det\left(\left[\hbar\braket{\mu_{a}|e^{i\zeta_{1}\hat{x}}\frac{1}{2\cosh\frac{\hat{p}}{2}}e^{i\zeta_{2}\hat{x}}\left(\frac{1}{2\cosh\frac{\hat{x}}{2}}\right)^{F}|\nu_{b}}\right]_{a,b}^{N\times N}\right)\nonumber \\
 & \quad\times\det\left(\left[\hbar\braket{\nu_{a}|\frac{1}{2\cosh\frac{\hat{p}}{2}}|\rho_{b}}\right]_{a,b}^{N\times N}\right)\nonumber \\
 & \quad\times\det\left(\left[\hbar\braket{\rho_{a}|e^{i\zeta_{3}\hat{x}}\frac{1}{2\cosh\frac{\hat{p}}{2}}e^{i\zeta_{4}\hat{x}}|\sigma_{b}}\right]_{a,b}^{N\times N}\right)\nonumber \\
 & \quad\times\det\left(\left[\hbar\braket{\sigma_{a}|\frac{1}{2\cosh\frac{\hat{p}}{2}}|\mu_{b}}\right]_{a,b}^{N\times N}\right).
\end{align}
Because in this case there are no rank deformations, we can directly apply \eqref{eq:Glue-Formula}. We then finally arrive at
\begin{equation}
Z_{\bm{k},\bm{N},F}^{\hat{A}_{3}}\left(\bm{\zeta},m\right)=\int\frac{d^{N}\mu}{N!}\det\left(\left[\braket{\mu_{a}|\hat{\rho}_{0,0,F}^{\hat{A}_{3}}\left(\hat{x},\hat{p};\bm{\zeta}\right)|\mu_{b}}\right]_{a,b}^{N\times N}\right),\label{eq:PF-A3-FGF2}
\end{equation}
where the density matrix is defined by
\begin{align}
\hat{\rho}_{0,0,F}^{\hat{A}_{3}}\left(\hat{x},\hat{p};\bm{\zeta}\right) & =e^{i\zeta_{1}\hat{x}}\frac{1}{2\cosh\frac{\hat{p}}{2}}e^{i\zeta_{2}\hat{x}}\left(\frac{1}{2\cosh\frac{\hat{x}}{2}}\right)^{F}\frac{1}{2\cosh\frac{\hat{p}}{2}}e^{i\zeta_{3}\hat{x}}\frac{1}{2\cosh\frac{\hat{p}}{2}}e^{i\zeta_{4}\hat{x}}\frac{1}{2\cosh\frac{\hat{p}}{2}}.\label{eq:DM-A32}
\end{align}
Here $\bm{k}$ and $\bm{N}$ are defined in \eqref{eq:A3-para}.

Finally, we consider the grand partition function defined in \eqref{eq:GPF-A3-def}. According to the formula \eqref{eq:GPF-Circle}, we obtain
\begin{equation}
\Xi_{0,0,F}^{\hat{A}_{3}}\left(\kappa;\bm{\zeta}\right)={\rm Det}\left(1+\kappa\hat{\rho}_{0,0,F}^{\hat{A}_{3}}\left(\hat{x},\hat{p};\bm{\zeta}\right)\right).\label{eq:GPF-A3-FGF2}
\end{equation}

\subsection{$\hat{D}_{3}$ theory\label{subsec:FGF-D3}}

In this section we apply the Fermi gas formalism to the matrix model for the $\hat{D}_{3}$ theory. Although the $\hat{D}_{3}$ matrix model \eqref{eq:PF_D3} is the same with the $\hat{A}_{3}$ matrix model, the way of applying the Fermi gas formalism is different from the one. The Fermi gas formalism for the $\hat{D}_{3}$ matrix model without rank deformations has been studied in \cite{Assel:2015hsa,Moriyama:2015jsa}.\footnote{The Fermi gas formalism for the $\hat{D}_{3}$ matrix model with rank deformations has been also studied in \cite{Kubo:2024raz}. However, the assignment of the Chern-Simons levels is different, and the procedure of the Fermi gas formalism is sensitive to the assignment. In this sense, our procedure is new.} In this section, we develop this procedure to the case that the rank deformation is turned on.

In this paper we consider the case of
\begin{equation}
\left(k_{1},k_{2},k_{3},k_{4}\right)=\left(-k,k,-k,k\right),\quad\left(N_{1},N_{2},N_{3},N_{4}\right)=\left(N,N,N,N+M\right),\label{eq:D3-para}
\end{equation}
with $M\geq0$. This is the same with the $\hat{A}_{3}$ theory case \eqref{eq:A3-para} because we will use the equality between $\hat{A}_{3}$ and $\hat{D}_{3}$. For this parameterization, we introduce the grand partition function which is defined as
\begin{equation}
\Xi_{k,M,F}^{\hat{D}_{3}}\left(\kappa;\bm{\zeta}\right)=\sum_{N=0}^{\infty}\kappa^{N}Z_{\bm{k},\bm{N},F}^{\hat{D}_{3}}\left(\bm{\zeta}\right).\label{eq:GPF-D3-def}
\end{equation}

We start the computation. The important first step is to insert the identity
\begin{equation}
1=\frac{\prod_{a,b}^{N,N}2\sinh\frac{\alpha_{a}^{\left(1\right)}-\alpha_{b}^{\left(3\right)}}{2}}{\prod_{a,b}^{N,N}2\sinh\frac{\alpha_{a}^{\left(1\right)}-\alpha_{b}^{\left(3\right)}}{2}}\frac{\prod_{a,b}^{N,N+M}2\sinh\frac{\alpha_{a}^{\left(2\right)}-\alpha_{b}^{\left(4\right)}}{2}}{\prod_{a,b}^{N,N+M}2\sinh\frac{\alpha_{a}^{\left(2\right)}-\alpha_{b}^{\left(4\right)}}{2}}.
\end{equation}
We also change the integration variables and introduce combined variables
\begin{align}
 & \mu_{a}=\alpha_{a}^{\left(1\right)},\quad\mu_{a}'=\alpha_{a}^{\left(3\right)},\quad\nu_{a}=\alpha_{a}^{\left(2\right)},\quad\nu_{a}'=\alpha_{a}^{\left(4\right)},\nonumber \\
 & \bar{\mu}_{a}=\begin{cases}
\mu_{a} & \left(1\leq a\leq N\right)\\
\mu_{a-N}' & \left(N+1\leq a\leq2N\right)
\end{cases},\quad\bar{\nu}_{a}=\begin{cases}
\nu_{a} & \left(1\leq a\leq N\right)\\
\nu_{a-N}' & \left(N+1\leq a\leq2N+M\right)
\end{cases}.
\end{align}
The 1-loop determinant factor can be recombined into
\begin{align}
\prod_{r=1}^{4}\frac{\prod_{a<a'}^{N_{r}}\left(2\sinh\frac{\alpha_{a}^{\left(r\right)}-\alpha_{a'}^{\left(r\right)}}{2}\right)^{2}}{\prod_{a,b}^{N_{r},N_{r+1}}2\cosh\frac{\alpha_{a}^{\left(r\right)}-\alpha_{b}^{\left(r+1\right)}}{2}} & =\left(-1\right)^{N\left(N+M\right)}\frac{\prod_{a<a'}^{N}2\sinh\frac{\mu_{a}-\mu_{a'}}{2}\prod_{b<b'}^{N}2\sinh\frac{\mu_{b}'-\mu_{b'}'}{2}}{\prod_{a,b}^{N,N}2\sinh\frac{\mu_{a}-\mu_{b}'}{2}}\nonumber \\
 & \quad\times\frac{\prod_{a<a'}^{2N}2\sinh\frac{\bar{\mu}_{a}-\bar{\mu}_{a'}}{2}\prod_{b<b'}^{2N+M}2\sinh\frac{\bar{\nu}_{b}-\bar{\nu}_{b'}}{2}}{\prod_{a,b}^{2N,2N+M}2\cosh\frac{\bar{\mu}_{a}-\bar{\nu}_{b}}{2}}\nonumber \\
 & \quad\times\frac{\prod_{a<a'}^{N+M}2\sinh\frac{\nu_{a}'-\nu_{a'}'}{2}\prod_{b<b'}^{N}2\sinh\frac{\nu_{b}-\nu_{b'}}{2}}{\prod_{a,b}^{N+M,N}2\sinh\frac{\nu_{a}'-\nu_{b}}{2}}.\label{eq:PF-D-recomb}
\end{align}
In the following we restrict our attention to two cases which are the same with $\hat{A}_{3}$.

\subsubsection{$F=0$ case}

The first case is when $F=0$ and $k>0$. After recombining the 1-loop determinants as \eqref{eq:PF-D-recomb} and rescaling the integration variables to $\alpha\rightarrow\alpha/k$, we can apply the determinant formulas \eqref{eq:DetFormula-cosh} and \eqref{eq:DetFormula-sinh} with setting $\hbar=2\pi k$. After putting appropriately the remaining factors to the determinants, we obtain
\begin{align}
 & Z_{\bm{k},\bm{N},0}^{\hat{D}_{3}}\left(\bm{\zeta}\right)\nonumber \\
 & =\frac{i^{-MN-\frac{1}{2}M^{2}}}{\hbar^{4N+M}}\int\frac{d^{N}\mu}{N!}\frac{d^{N}\mu'}{N!}\frac{d^{N}\nu}{N!}\frac{d^{N+M}\nu'}{\left(N+M\right)!}\nonumber \\
 & \quad\times\det\left(\left[\hbar\braket{\mu_{a}|e^{\frac{i\zeta_{1}}{k}\hat{x}}e^{-\frac{i}{4\pi k}\hat{x}^{2}}\frac{1}{2}\tanh\frac{\hat{p}}{2}e^{-\frac{i}{4\pi k}\hat{x}^{2}}e^{\frac{i\zeta_{3}}{k}\hat{x}}|\mu_{b}'}\right]_{a,b}^{N\times N}\right)\nonumber \\
 & \quad\times\det\left(\begin{array}{c}
\left[\hbar\braket{\bar{\mu}_{a}|\frac{1}{2\cosh\frac{\hat{p}-i\pi M}{2}}|\bar{\nu}_{b}}\right]_{a,b}^{2N\times\left(2N+M\right)}\\
\left[\frac{\hbar}{\sqrt{k}}\bbraket{t_{M,j}|\bar{\nu}_{b}}\right]_{j,b}^{M\times\left(2N+M\right)}
\end{array}\right)\nonumber \\
 & \quad\times\det\left(\begin{array}{cc}
\left[\hbar\braket{\nu_{a}'|e^{\frac{i\zeta_{4}}{k}\hat{x}}e^{\frac{i}{4\pi k}\hat{x}^{2}}\frac{1}{2}\tanh\frac{\hat{p}-i\pi M}{2}e^{\frac{i}{4\pi k}\hat{x}^{2}}e^{\frac{i\zeta_{2}}{k}\hat{x}}|\nu_{b}}\right]_{a,b}^{\left(N+M\right)\times N} & \left[A_{a,j}\right]_{a,j}^{\left(N+M\right)\times M}\end{array}\right),
\end{align}
where
\begin{equation}
A_{a,j}=\frac{\hbar}{\sqrt{k}}\brakket{\nu_{a}'|e^{\frac{i\zeta_{4}}{k}\hat{x}}e^{\frac{i}{2\hbar}\hat{x}^{2}}|-t_{M,j}}.
\end{equation}
The third determinant can be diagonalized by using the formula \eqref{eq:DiagFormula}. 

For generating the delta functions for performing the integration over $\nu'_{n+j}$ ($j=1,2,\ldots,M$), we perform similarity transformations as well as the ${\rm U}\times{\rm U}$ and $\hat{A}_{3}$ matrix model. However, this time we need a trick for $\mu$ and $\nu$ because the form of the vectors of $\mu$ or $\nu$ is not $\ket{\alpha}\bra{\alpha}$ but $\bra{\mu}\bra{\mu}$ or $\ket{\nu}\ket{\nu}$. We can make the form of $\ket{\alpha}\bra{\alpha}$ by using the transpose operation. The properties of it is summarized in appendix \ref{sec:QM}. For $\mu$, the similarity transformations can be performed as
\begin{align}
 & \int d\mu\braket{\mu|\hat{{\cal O}}_{1}|\mu'}\bra{\mu}\hat{{\cal O}}_{2}=\int d\mu\braket{\mu'|\hat{{\cal O}}_{1}^{t}|\mu}\bra{\mu}\hat{{\cal O}}_{2}\nonumber \\
 & =\int d\mu\braket{\mu'|\hat{{\cal O}}_{1}^{t}\hat{{\cal O}}_{2}|\mu}\bra{\mu}=\int d\mu\braket{\mu|\hat{{\cal O}}_{2}^{t}\hat{{\cal O}}_{1}|\mu'}\bra{\mu}.
\end{align}
In this manner, we can perform the similarity transformations
\begin{align}
 & \int d\mu\bra{\mu}\bra{\mu}\rightarrow\int d\mu\bra{\mu}e^{-\frac{i}{2\hbar}\hat{p}^{2}}\bra{\mu}e^{\frac{i}{2\hbar}\hat{p}^{2}},\quad\int d\mu'\ket{\mu'}\bra{\mu'}=\int d\mu'e^{-\frac{i}{2\hbar}\hat{p}^{2}}\ket{\mu'}\bra{\mu'}e^{\frac{i}{2\hbar}\hat{p}^{2}},\nonumber \\
 & \int d\nu\ket{\nu}\ket{\nu}\rightarrow\int d\nu e^{-\frac{i}{2\hbar}\hat{p}^{2}}\ket{\nu}e^{\frac{i}{2\hbar}\hat{p}^{2}}\ket{\nu},\quad\int d\nu'\ket{\nu'}\bra{\nu'}=\int d\nu'e^{-\frac{i}{2\hbar}\hat{p}^{2}}\ket{\nu'}\bra{\nu'}e^{\frac{i}{2\hbar}\hat{p}^{2}}.
\end{align}
Here, $\bra{\mu}e^{-\frac{i}{2\hbar}\hat{p}^{2}}$ is in the first determinant while $\bra{\mu}e^{\frac{i}{2\hbar}\hat{p}^{2}}$ is in the second determinant. Similarly, $e^{-\frac{i}{2\hbar}\hat{p}^{2}}\ket{\nu}$ is in the second determinant while $e^{\frac{i}{2\hbar}\hat{p}^{2}}\ket{\nu}$ is in the third determinant. After performing this similarity transformation, $A_{a,j}$ becomes
\begin{equation}
A_{a,j}\rightarrow\frac{\hbar}{\sqrt{k}}\brakket{\nu_{a}'|e^{\frac{i}{4\pi k}\hat{p}^{2}}e^{\frac{i\zeta_{4}}{k}\hat{x}}e^{\frac{i}{4\pi k}\hat{x}^{2}}|-t_{M,j}}.
\end{equation}
Thus, by using \eqref{eq:VecSim}, $M$ delta functions $\delta\left(\nu_{N+j}'+2\pi\zeta_{4}-t_{M,j}\right)$ ($j=1,2,\ldots,M$) appear on the third determinant.

After applying the determinant formula \eqref{eq:DetFormula-cosh} backwards to the second determinant, we can perform the integration over $\nu_{N+j}'$ ($j=1,2,\ldots,M$) by using the delta functions. We apply the determinant formula \eqref{eq:DetFormula-cosh} again to the factor coming from the second determinant and put the remaining factors to the determinant appropriately. We also apply the diagonalizing formula \eqref{eq:DiagFormula} backwards. At this stage we have
\begin{align}
 & Z_{\bm{k},\bm{N},0}^{\hat{D}_{3}}\left(\bm{\zeta}\right)\nonumber \\
 & =Z_{k,M}^{{\rm U}}\left(\zeta_{4}\right)\frac{i^{-MN}}{2^{2N}}\int\frac{d^{N}\mu}{N!}\frac{d^{N}\mu'}{N!}\frac{d^{N}\nu}{N!}\frac{d^{N}\nu'}{N!}\nonumber \\
 & \quad\times\det\left(\left[\braket{\mu_{a}|e^{-\frac{i}{4\pi k}\hat{p}^{2}}e^{\frac{i\zeta_{1}}{k}\hat{x}}e^{-\frac{i}{4\pi k}\hat{x}^{2}}\tanh\frac{\hat{p}}{2}e^{-\frac{i}{4\pi k}\hat{x}^{2}}e^{\frac{i\zeta_{3}}{k}\hat{x}}e^{-\frac{i}{4\pi k}\hat{p}^{2}}|\mu_{b}'}\right]_{a,b}^{N\times N}\right)\nonumber \\
 & \quad\times\det\left(\left[\braket{\bar{\mu}_{a}|\frac{1}{\prod_{j}^{M}2\cosh\frac{\hat{x}+2\pi\zeta_{4}-t_{M,j}}{2k}}\frac{1}{2\cosh\frac{\hat{p}}{2}}\prod_{j}^{M}2\sinh\frac{\hat{x}+2\pi\zeta_{4}-t_{M,j}}{2k}|\bar{\nu}_{b}}\right]_{a,b}^{2N\times2N}\right)\nonumber \\
 & \quad\times\det\left(\left[\braket{\nu_{a}'|e^{\frac{i}{4\pi k}\hat{p}^{2}}e^{\frac{i\zeta_{4}}{k}\hat{x}}e^{\frac{i}{4\pi k}\hat{x}^{2}}\tanh\frac{\hat{p}-i\pi M}{2}e^{\frac{i}{4\pi k}\hat{x}^{2}}e^{\frac{i\zeta_{2}}{k}\hat{x}}e^{\frac{i}{4\pi k}\hat{p}^{2}}|\nu_{b}}\right]_{a,b}^{N\times N}\right),
\end{align}
where $Z_{k,M}^{{\rm U}}$ is defined in \eqref{eq:Z_U-Def}.

Finally, we consider the grand partition function defined in \eqref{eq:GPF-D3-def}. According to the formula \eqref{eq:GPF-Orient}, we obtain
\begin{equation}
\Xi_{k,M,0}^{\hat{D}_{3}}\left(\kappa;\bm{\zeta}\right)=Z_{k,M}^{{\rm U}}\left(\zeta_{4}\right)\sqrt{{\rm Det}\left(1+\kappa\hat{\rho}_{k,M,0}^{\hat{D}_{3}}\left(\hat{x},\hat{p};\bm{\zeta}\right)\right)},\label{eq:GPF-D3-FGF1}
\end{equation}
where the density matrix is defined by
\begin{align}
\hat{\rho}_{k,M,0}^{\hat{D}_{3}}\left(\hat{x},\hat{p};\bm{\zeta}\right) & =\frac{i^{-M}}{4}e^{-\frac{i}{4\pi k}\hat{p}^{2}}e^{-\frac{i}{4\pi k}\hat{x}^{2}}\left(e^{\frac{i\zeta_{1}}{k}\hat{x}}\tanh\frac{\hat{p}}{2}e^{\frac{i\zeta_{3}}{k}\hat{x}}+e^{\frac{i\zeta_{3}}{k}\hat{x}}\tanh\frac{\hat{p}}{2}e^{\frac{i\zeta_{1}}{k}\hat{x}}\right)e^{-\frac{i}{4\pi k}\hat{x}^{2}}e^{-\frac{i}{4\pi k}\hat{p}^{2}}\nonumber \\
 & \quad\times\frac{1}{\prod_{j}^{M}2\cosh\frac{\hat{x}+2\pi\zeta_{4}-t_{M,j}}{2k}}\frac{1}{2\cosh\frac{\hat{p}}{2}}\prod_{j}^{M}2\sinh\frac{\hat{x}+2\pi\zeta_{4}-t_{M,j}}{2k}\nonumber \\
 & \quad\times e^{\frac{i}{4\pi k}\hat{p}^{2}}e^{\frac{i}{4\pi k}\hat{x}^{2}}\left(e^{\frac{i\zeta_{4}}{k}\hat{x}}\tanh\frac{\hat{p}-i\pi M}{2}e^{\frac{i\zeta_{2}}{k}\hat{x}}+e^{\frac{i\zeta_{2}}{k}\hat{x}}\tanh\frac{\hat{p}+i\pi M}{2}e^{\frac{i\zeta_{4}}{k}\hat{x}}\right)e^{\frac{i}{4\pi k}\hat{x}^{2}}e^{\frac{i}{4\pi k}\hat{p}^{2}}\nonumber \\
 & \quad\times\prod_{j}^{M}2\sinh\frac{\hat{x}+2\pi\zeta_{4}-t_{M,j}}{2k}\frac{1}{2\cosh\frac{\hat{p}}{2}}\frac{1}{\prod_{j}^{M}2\cosh\frac{\hat{x}+2\pi\zeta_{4}-t_{M,j}}{2k}}.\label{eq:DM-D31}
\end{align}

\subsubsection{$k=0$ and $M=0$ case}

The second case is when $k=0$ and $M=0$. After recombining the 1-loop determinants as \eqref{eq:PF-D-recomb}, we can apply the determinant formulas \eqref{eq:DetFormula-cosh} and \eqref{eq:DetFormula-sinh} with setting $\hbar=2\pi$. After putting appropriately the remaining factors to the determinants, we obtain
\begin{align}
 & Z_{\bm{k},\bm{N},F}^{\hat{D}_{3}}\left(\bm{\zeta}\right)\nonumber \\
 & =\frac{1}{2^{2N}}\int\frac{d^{N}\mu}{N!}\frac{d^{N}\mu'}{N!}\frac{d^{N}\nu}{N!}\frac{d^{N}\nu'}{N!}\det\left(\left[\braket{\mu_{a}|e^{\frac{i\zeta_{1}}{k}\hat{x}}\tanh\frac{\hat{p}}{2}e^{\frac{i\zeta_{3}}{k}\hat{x}}|\mu_{b}'}\right]_{a,b}^{N\times N}\right)\nonumber \\
 & \quad\times\det\left(\left[\braket{\bar{\mu}_{a}|\frac{1}{2\cosh\frac{\hat{p}}{2}}|\bar{\nu}_{b}}\right]_{a,b}^{2N\times2N}\right)\det\left(\left[\braket{\nu_{a}'|e^{\frac{i\zeta_{4}}{k}\hat{x}}\tanh\frac{\hat{p}}{2}\frac{e^{\frac{i\zeta_{2}}{k}\hat{x}}}{\left(2\cosh\frac{\hat{x}}{2}\right)^{F}}|\nu_{b}}\right]_{a,b}^{N\times N}\right).
\end{align}

Finally, we consider the grand partition function defined in \eqref{eq:GPF-D3-def}. According to the formula \eqref{eq:GPF-Orient}, we obtain
\begin{equation}
\Xi_{0,0,F}^{\hat{D}_{3}}\left(\kappa;\bm{\zeta}\right)=\sqrt{{\rm Det}\left(1+\kappa\hat{\rho}_{0,0,F}^{\hat{D}_{3}}\left(\hat{x},\hat{p};\bm{\zeta}\right)\right)},\label{eq:GPF-D3-FGF2}
\end{equation}
where the density matrix is defined by
\begin{align}
\hat{\rho}_{0,0,F}^{\hat{D}_{3}}\left(\hat{x},\hat{p};\bm{\zeta}\right) & =\frac{1}{4}\left(e^{\frac{i\zeta_{1}}{k}\hat{x}}\tanh\frac{\hat{p}}{2}e^{\frac{i\zeta_{3}}{k}\hat{x}}+e^{\frac{i\zeta_{3}}{k}\hat{x}}\tanh\frac{\hat{p}}{2}e^{\frac{i\zeta_{1}}{k}\hat{x}}\right)\frac{1}{2\cosh\frac{\hat{p}}{2}}\nonumber \\
 & \quad\times\left(e^{\frac{i\zeta_{4}}{k}\hat{x}}\tanh\frac{\hat{p}}{2}\frac{e^{\frac{i\zeta_{2}}{k}\hat{x}}}{\left(2\cosh\frac{\hat{x}}{2}\right)^{F}}+\frac{e^{\frac{i\zeta_{2}}{k}\hat{x}}}{\left(2\cosh\frac{\hat{x}}{2}\right)^{F}}\tanh\frac{\hat{p}}{2}e^{\frac{i\zeta_{4}}{k}\hat{x}}\right)\frac{1}{2\cosh\frac{\hat{p}}{2}}.\label{eq:DM-D32}
\end{align}

\subsection{${\rm O}\left(2N+M\right)_{2k}\times{\rm USp}\left(2N\right)_{-k}$ theory\label{subsec:FGF-OUSp}}

In this section we apply the Fermi gas formalism for the matrix model for the ${\rm O}\left(2N+M\right)_{2k}\times{\rm USp}\left(2N\right)_{-k}$ theory. The Fermi gas approach for ${\rm O}\left(2N+M\right)_{2k}\times{\rm USp}\left(2N\right)_{-k}$ matrix model had been studied in \cite{Moriyama:2015asx,Moriyama:2016kqi,Moriyama:2016xin}. In this section we suggest a different Fermi gas approach which is suitable for the mass deformation. This new approach, however, is not applicable for $M\geq2$ at the moment. Also, the matrix model for odd and even $M$ are different. For this reason, in the following we consider $M=0$ with mass deformations, $M=1$ with mass deformations and $M=2$ without mass deformation separately. We leave $M\geq2$ with mass deformations to future work.

Before going to the detail, we introduce the grand partition function for general $M\geq0$
\begin{equation}
\Xi_{k,M}^{{\rm O}\times{\rm USp}}\left(\kappa;m\right)=\sum_{N=0}^{\infty}\kappa^{N}Z_{k,\left(2N+M,2N\right)}^{{\rm O}\times{\rm USp}}\left(m\right).\label{eq:GPF-OUSp-def}
\end{equation}

\subsubsection{$M=0$ case\label{subsec:FGF-OUSp-M0}}

We start with the ${\rm O}\left(2N\right)_{2k}\times{\rm USp}\left(2N\right)_{-k}$ matrix model \eqref{eq:PF_OUSpE} with rescaling the integration variables as $\alpha\rightarrow\alpha/k$. First, we introduce new variables 
\begin{align}
 & \mu_{a}'=-\mu_{a},\quad\nu_{a}'=-\nu_{a},\nonumber \\
 & \bar{\mu}_{a}=\begin{cases}
\mu_{a} & \left(1\leq a\leq N\right)\\
\mu_{a-N}' & \left(N+1\leq a\leq2N\right)
\end{cases},\quad\bar{\nu}_{a}=\begin{cases}
\nu_{a} & \left(1\leq a\leq N\right)\\
\nu_{a-N}' & \left(N+1\leq a\leq2N\right)
\end{cases}.
\end{align}
The first line is realized by inserting
\begin{align}
 & 1=\int d^{N}\mu'\int d^{N}\nu'\prod_{a=1}^{N}\braket{\mu_{a}|\hat{R}|\mu_{a}'}\prod_{b=1}^{N}\braket{\nu_{b}'|\hat{R}|\nu_{b}}.
\end{align}
The definition and the property of the reflection operator $\hat{R}$ is summarized in appendix \ref{sec:QM}. The 1-loop determinant is reformulated into
\begin{align}
 & \frac{\prod_{a<a'}^{N}\left(2\sinh\frac{\mu_{a}-\mu_{a'}}{2}2\sinh\frac{\mu_{a}+\mu_{a'}}{2}\right)^{2}\prod_{b<b'}^{N}\left(2\sinh\frac{\nu_{b}-\nu_{b'}}{2}2\sinh\frac{\nu_{b}+\nu_{b'}}{2}\right)^{2}}{\prod_{a,b}^{N,N}\left(2\cosh\frac{\mu_{a}-\nu_{b}+2\pi m}{2}\right)\left(2\cosh\frac{\mu_{a}+\nu_{b}+2\pi m}{2}\right)\left(2\cosh\frac{\mu_{a}-\nu_{b}-2\pi m}{2}\right)\left(2\cosh\frac{\mu_{a}+\nu_{b}-2\pi m}{2}\right)}\nonumber \\
 & =\frac{1}{\prod_{a}^{N}2\sinh\mu_{a}\prod_{b}^{N}2\sinh\nu_{b}}\frac{\prod_{a<a'}^{2N}2\sinh\frac{\bar{\mu}_{a}-\bar{\mu}_{a'}}{2}\prod_{b<b'}^{2N}2\sinh\frac{\bar{\nu}_{b}-\bar{\nu}_{b'}}{2}}{\prod_{a,b}^{2N,2N}2\cosh\frac{\bar{\mu}_{a}-\bar{\nu}_{b}+2\pi m}{2}}.
\end{align}
By using \eqref{eq:DetFormula-cosh} with setting $\hbar=2\pi k$, the last factor can be written by a determinant of a matrix. By using \eqref{eq:DiagFormula} backwards, we can also write the delta functions as a determinant of a matrix
\begin{align}
 & \int d^{N}\mu'\int d^{N}\nu'\prod_{a=1}^{N}\braket{\mu_{a}|\hat{R}|\mu_{a}'}\prod_{b=1}^{N}\braket{\nu_{b}'|\hat{R}|\nu_{b}}\nonumber \\
 & \rightarrow\int\frac{d^{N}\mu'}{N!}\int\frac{d^{N}\nu'}{N!}\det\left(\left[\braket{\mu_{a}|\hat{R}|\mu_{b}'}\right]_{a,b}^{N\times N}\right)\det\left(\left[\braket{\nu_{a}'|\hat{R}|\nu_{b}}\right]_{a,b}^{N\times N}\right).
\end{align}
After putting the Chern-Simons factors and remaining factors into the determinants appropriately, we arrive at
\begin{align}
 & Z_{k,\left(2N,2N\right)}^{{\rm O}\times{\rm USp}}\left(m\right)\nonumber \\
 & =\frac{1}{2^{2N}}\int\frac{d^{N}\mu}{N!}\frac{d^{N}\nu}{N!}\frac{d^{N}\mu'}{N!}\frac{d^{N}\nu'}{N!}\det\left(\left[\braket{\mu_{a}|\frac{i}{\sinh\hat{x}}\hat{R}|\mu_{b}'}\right]_{a,b}^{N\times N}\right)\nonumber \\
 & \quad\times\det\left(\left[\braket{\bar{\mu}_{a}|e^{\frac{i}{4\pi k}\hat{x}^{2}}\frac{e^{im\hat{p}}}{2\cosh\frac{\hat{p}}{2}}e^{-\frac{i}{4\pi k}\hat{x}^{2}}|\bar{\nu}_{b}}\right]_{a,b}^{2N\times2N}\right)\det\left(\left[\braket{\nu_{a}'|\hat{R}\sinh\hat{x}|\nu_{b}}\right]_{a,b}^{N\times N}\right).
\end{align}

Finally, we consider the grand partition function defined in \eqref{eq:GPF-OUSp-def}. According to the formula \eqref{eq:GPF-Orient}, we obtain
\begin{equation}
\Xi_{k,0}^{{\rm O}\times{\rm USp}}\left(\kappa;m\right)=\sqrt{{\rm Det}\left(1+\kappa\hat{\rho}_{k,0}^{{\rm O}\times{\rm USp}}\left(\hat{x},\hat{p};m\right)\right)},\label{eq:GPF-OUSpM0-FGF}
\end{equation}
where the density matrix is defined by
\begin{align}
\hat{\rho}_{k,0}^{{\rm O}\times{\rm USp}}\left(\hat{x},\hat{p};m\right) & =\frac{i}{2\sinh\hat{x}}\hat{R}e^{\frac{i}{4\pi k}\hat{x}^{2}}\frac{e^{im\hat{p}}}{2\cosh\frac{\hat{p}}{2}}e^{-\frac{i}{4\pi k}\hat{x}^{2}}\hat{R}\left(2\sinh\hat{x}\right)e^{-\frac{i}{4\pi k}\hat{x}^{2}}\frac{e^{-im\hat{p}}}{2\cosh\frac{\hat{p}}{2}}e^{\frac{i}{4\pi k}\hat{x}^{2}}.\label{eq:DM-OUSpM0}
\end{align}

\subsubsection{$M=1$ case\label{subsec:FGF-OUSp-M1}}

In this section we focus on the $M=1$ case. We start with the ${\rm O}\left(2N+1\right)_{2k}\times{\rm USp}\left(2N\right)_{-k}$ matrix model \eqref{eq:PF_OUSpO} with rescaling the integration variables as $\alpha\rightarrow\alpha/k$. The flow of the computation is almost the same with the ${\rm O}\left(2N\right)_{2k}\times{\rm USp}\left(2N\right)_{-k}$ case.

First, we introduce new variables
\begin{align}
 & \mu_{a}'=-\mu_{a},\quad\nu_{a}'=-\nu_{a},\nonumber \\
 & \bar{\mu}_{a}=\begin{cases}
\mu_{a} & \left(1\leq a\leq N\right)\\
\mu_{a-N}' & \left(N+1\leq a\leq2N\right)
\end{cases},\quad\bar{\nu}_{a}=\begin{cases}
\nu_{a} & \left(1\leq a\leq N\right)\\
\nu_{a-N}' & \left(N+1\leq a\leq2N\right)
\end{cases}.
\end{align}
The first line is realized by inserting
\begin{align}
 & 1=\int d^{N}\mu'\int d^{N}\nu'\prod_{a=1}^{N}\braket{\mu_{a}|\hat{R}|\mu_{a}'}\prod_{a=1}^{N}\braket{\nu_{a}'|\hat{R}|\nu_{a}}.
\end{align}
The definition and the property of the reflection operator $\hat{R}$ is summarized in appendix \ref{sec:QM}. The 1-loop determinant is reformulated into
\begin{align}
 & \frac{\prod_{a<a'}^{N}\left(2\sinh\frac{\mu_{a}-\mu_{a'}}{2}2\sinh\frac{\mu_{a}+\mu_{a'}}{2}\right)^{2}\prod_{b<b'}^{N}\left(2\sinh\frac{\nu_{b}-\nu_{b'}}{2}2\sinh\frac{\nu_{b}+\nu_{b'}}{2}\right)^{2}}{\prod_{a,b}^{N,N}\left(2\cosh\frac{\mu_{a}-\nu_{b}+2\pi m}{2}\right)\left(2\cosh\frac{\mu_{a}+\nu_{b}+2\pi m}{2}\right)\left(2\cosh\frac{\mu_{a}-\nu_{b}-2\pi m}{2}\right)\left(2\cosh\frac{\mu_{a}+\nu_{b}-2\pi m}{2}\right)}\nonumber \\
 & =\frac{1}{\prod_{a}^{N}2\sinh\mu_{a}\prod_{b}^{N}2\sinh\nu_{b}}\frac{\prod_{a<a'}^{2N}2\sinh\frac{\bar{\mu}_{a}-\bar{\mu}_{a'}}{2}\prod_{b<b'}^{2N}2\sinh\frac{\bar{\nu}_{b}-\bar{\nu}_{b'}}{2}}{\prod_{a,b}^{2N,2N}2\cosh\frac{\bar{\mu}_{a}-\bar{\nu}_{b}+2\pi m}{2}}.
\end{align}
By using \eqref{eq:DetFormula-cosh} with setting $\hbar=2\pi k$, the last factor can be written by a determinant of a matrix. By using \eqref{eq:DiagFormula} backwards, we can also write the delta functions as a determinant of a matrix
\begin{align}
 & \int d^{N}\mu'\int d^{N}\nu'\prod_{a=1}^{N}\braket{\mu_{a}|\hat{R}|\mu_{a}'}\prod_{b=1}^{N}\braket{\nu_{b}'|\hat{R}|\nu_{b}}\nonumber \\
 & \rightarrow\int\frac{d^{N}\mu'}{N!}\int\frac{d^{N}\nu'}{N!}\det\left(\left[\braket{\mu_{a}|\hat{R}|\mu_{b}'}\right]_{a,b}^{N\times N}\right)\det\left(\left[\braket{\nu_{a}'|\hat{R}|\nu_{b}}\right]_{a,b}^{N\times N}\right).
\end{align}
After putting the Chern-Simons factors and remaining factors into the determinants appropriately, we arrive at
\begin{align}
Z_{k,\left(2N+1,2N\right)}^{{\rm O}\times{\rm USp}}\left(m\right) & =\frac{1}{2^{2N+1}}\int\frac{d^{N}\mu}{N!}\frac{d^{N}\nu}{N!}\frac{d^{N}\mu'}{N!}\frac{d^{N}\nu'}{N!}\det\left(\left[\braket{\mu_{a}|\tanh\frac{\hat{x}}{2}\hat{R}|\mu_{b}'}\right]_{a,b}^{N\times N}\right)\nonumber \\
 & \quad\times\det\left(\left[\braket{\bar{\mu}_{a}|e^{\frac{i}{4\pi k}\hat{x}^{2}}\frac{e^{im\hat{p}}}{2\cosh\frac{\hat{p}}{2}}e^{-\frac{i}{4\pi k}\hat{x}^{2}}|\bar{\nu}_{b}}\right]_{a,b}^{2N\times2N}\right)\nonumber \\
 & \quad\times\det\left(\left[\braket{\nu_{a}'|\hat{R}\frac{2\sinh\hat{x}}{\left(2\cosh\frac{\hat{x}+2\pi m}{2}\right)\left(2\cosh\frac{\hat{x}-2\pi m}{2}\right)}|\nu_{b}}\right]_{a,b}^{N\times N}\right).
\end{align}

Finally, we consider the grand partition function defined in \eqref{eq:GPF-OUSp-def}. According to the formula \eqref{eq:GPF-Orient}, we obtain
\begin{equation}
\Xi_{k,1}^{{\rm O}\times{\rm USp}}\left(\kappa;m\right)=Z_{1}^{{\rm O}}\sqrt{{\rm Det}\left(1+\kappa\hat{\rho}_{k,1}^{{\rm O}\times{\rm USp}}\left(\hat{x},\hat{p};m\right)\right)},\label{eq:GPF-OUSpM1-FGF}
\end{equation}
where the density matrix is defined by
\begin{align}
\hat{\rho}_{k,1}^{{\rm O}\times{\rm USp}}\left(\hat{x},\hat{p};m\right) & =\tanh\frac{\hat{x}}{2}\hat{R}e^{\frac{i}{4\pi k}\hat{x}^{2}}\frac{e^{im\hat{p}}}{2\cosh\frac{\hat{p}}{2}}e^{-\frac{i}{4\pi k}\hat{x}^{2}}\nonumber \\
 & \quad\times\hat{R}\frac{2\sinh\hat{x}}{\left(2\cosh\frac{\hat{x}+2\pi m}{2}\right)\left(2\cosh\frac{\hat{x}-2\pi m}{2}\right)}e^{-\frac{i}{4\pi k}\hat{x}^{2}}\frac{e^{-im\hat{p}}}{2\cosh\frac{\hat{p}}{2}}e^{\frac{i}{4\pi k}\hat{x}^{2}}.\label{eq:DM-OUSpM1}
\end{align}
The overall factor $Z_{1}^{{\rm O}}$ would possibly be regarded as the partition function of the ${\rm O}\left(1\right)$ gauge theory, whose value is
\begin{equation}
Z_{1}^{{\rm O}}=\frac{1}{2}.\label{eq:Z_O1-Def}
\end{equation}

\subsubsection{$M=2$ case\label{subsec:FGF-OUSp-M2}}

In this section we apply the Fermi gas approach to ${\rm O}\left(2N+2\right)_{2k}\times{\rm USp}\left(2N\right)_{-k}$ matrix model without mass deformation. We start with \eqref{eq:PF_OUSpE} with $m=0$.

It was discovered in \cite{Moriyama:2016kqi} that the ${\rm O}\left(2N+2\right)_{2k}\times{\rm USp}\left(2N\right)_{-k}$ matrix model can be written as
\begin{equation}
Z_{k,\left(2N+2,2N\right)}^{{\rm O}\times{\rm USp}}\left(0\right)=Z_{2k,2}^{{\rm O}}\int\frac{d^{N}\mu}{N!}\det\left(\left[\braket{\mu_{a}|\hat{\rho}_{2k,1}^{{\rm U}\times{\rm U}}\left(\hat{x},\hat{p};0\right)\frac{1-\hat{R}}{2}|\mu_{b}}\right]_{a,b}^{N\times N}\right),
\end{equation}
where $Z_{k,2M}^{{\rm O}}$ is the $S^{3}$ partition function of the pure Chern-Simons theory with ${\rm O}\left(2M\right)_{k}$ gauge group defined by
\begin{equation}
Z_{k,2M}^{{\rm O}}=e^{\frac{\pi i}{3k}M\left(M-1\right)\left(2M-1\right)}\frac{1}{2k^{\frac{M}{2}}}\prod_{j<j'}^{M}4\sin\frac{\pi}{2k}\left(j'-j\right)\sin\frac{\pi}{2k}\left(j'+j-2\right).\label{eq:Z_O-Def}
\end{equation}

According to the formula \eqref{eq:GPF-CP}, the grand partition function defined in \eqref{eq:GPF-OUSp-def} becomes
\begin{equation}
\Xi_{k,2}^{{\rm O}\times{\rm USp}}\left(\kappa;0\right)=Z_{2k,2}^{{\rm O}}{\rm Det}\left(1+\kappa\hat{\rho}_{2k,1}^{{\rm U}\times{\rm U}}\left(\hat{x},\hat{p};0\right)\frac{1-\hat{R}}{2}\right).
\end{equation}
The ${\rm U}\times{\rm U}$ density matrix \eqref{eq:DM-UU} appeared here is 
\begin{equation}
\hat{\rho}_{2k,1}^{{\rm U}\times{\rm U}}\left(\hat{X},\hat{P};0\right)=\frac{2\sinh\frac{\hat{X}}{4k}}{2\sinh\frac{\hat{X}}{2}}\frac{1}{2\cosh\frac{\hat{P}}{2}}\frac{1}{2\cosh\frac{\hat{X}}{4k}}.
\end{equation}
We want to use the Fredholm determinant formula \eqref{eq:GPF-CP}. For that, we need to rewrite the density matrix which satisfies the conditions \eqref{eq:GPF-DMcond}. We introduce
\begin{equation}
\hat{x}=\frac{\hat{X}}{2},\quad\hat{p}=\hat{P}.
\end{equation}
Note that $\left[\hat{x},\hat{p}\right]=2\pi ik$. We use the property of the chiral projection
\begin{equation}
\frac{1}{\tanh\frac{\hat{x}}{2k}}\frac{1-\hat{R}}{4\cosh\frac{\hat{p}}{2}}=\frac{1+\hat{R}}{4\cosh\frac{\hat{p}}{2}}\tanh\frac{\hat{x}}{2k}=\frac{1}{2\cosh\frac{\hat{p}}{2}}\tanh\frac{\hat{x}}{2k}\frac{1-\hat{R}}{2}.
\end{equation}
One can check the first equality by putting them between position operators, and the second equality comes from the property of the reflection operator. By using this relation, one can show that
\begin{equation}
\frac{1-\hat{R}}{4\cosh\frac{\hat{p}}{2}}=\tanh\frac{\hat{x}}{k}\left(\frac{1}{\tanh\frac{\hat{x}}{2k}}+\tanh\frac{\hat{x}}{2k}\right)\frac{1-\hat{R}}{4\cosh\frac{\hat{p}}{2}}=\tanh\frac{\hat{x}}{k}\left\{ \tanh\frac{\hat{x}}{2k},\frac{1}{2\cosh\frac{\hat{p}}{2}}\right\} \frac{1-\hat{R}}{2}.
\end{equation}
The first equality comes from the property of the hyperbolic function. Thus, the density matrix can be rewritten as
\begin{align}
\hat{\rho}_{2k,1}^{{\rm U}\times{\rm U}}\left(2\hat{x},\hat{p};0\right)\frac{1-\hat{R}}{2} & \sim\hat{\rho}_{k,2}^{{\rm O}\times{\rm USp}}\left(\hat{x},\hat{p};0\right)\frac{1-\hat{R}}{2},
\end{align}
where
\begin{equation}
\hat{\rho}_{k,2}^{{\rm O}\times{\rm USp}}\left(\hat{x},\hat{p};0\right)=\frac{\tanh\frac{\hat{x}}{2k}}{2\sinh\hat{x}}\tanh\frac{\hat{x}}{k}\left\{ \tanh\frac{\hat{x}}{2k},\frac{1}{2\cosh\frac{\hat{p}}{2}}\right\} .\label{eq:DM-OUSpM2}
\end{equation}
Now $\hat{\rho}_{k,2}^{{\rm O}\times{\rm USp}}\left(\hat{x},\hat{p};0\right)$ satisfies the conditions \eqref{eq:GPF-DMcond} (see blow). Hence, according to the formula \eqref{eq:GPF-CP}, we obtain
\begin{equation}
\Xi_{k,2}^{{\rm O}\times{\rm USp}}\left(\kappa;0\right)=Z_{2k,2}^{{\rm O}}\sqrt{{\rm Det}\left(1+\kappa\hat{\rho}_{k,2}^{{\rm O}\times{\rm USp}}\left(\hat{x},\hat{p};0\right)\right)}.\label{eq:GPF-OUSpM2-FGF}
\end{equation}

We finally show that $\hat{\rho}_{k,2}^{{\rm O}\times{\rm USp}}\left(\hat{x},\hat{p};0\right)$ satisfies the conditions \eqref{eq:GPF-DMcond}. It is easy to show that the second condition is satisfied. To show the first condition, we decompose the density matrix as
\begin{equation}
\hat{\rho}_{k,2}^{{\rm O}\times{\rm USp}}\left(\hat{x},\hat{p};0\right)=\hat{f}_{1}\hat{f}_{2},
\end{equation}
where
\begin{equation}
\hat{f}_{1}=\frac{\tanh\frac{\hat{x}}{2k}}{2\sinh\hat{x}}\tanh\frac{\hat{x}}{k},\quad\hat{f}_{2}=\left\{ \tanh\frac{\hat{x}}{2k},\frac{1}{2\cosh\frac{\hat{p}}{2}}\right\} .
\end{equation}
One can easily check that both $\hat{f}_{1}$ and $\hat{f}_{2}$ satisfy the following two conditions
\begin{equation}
\hat{f}_{i}^{t}=\hat{f}_{i},\quad\left\{ \hat{f}_{i},\hat{R}\right\} =0.
\end{equation}
Hence
\begin{align}
 & {\rm tr}\left[\left(\hat{\rho}_{k,2}^{{\rm O}\times{\rm USp}}\left(\hat{x},\hat{p};0\right)\right)^{n}\hat{R}\right]={\rm tr}\left[\hat{R}^{t}\left(\hat{f}_{2}^{t}\hat{f}_{1}^{t}\right)^{n}\right]\nonumber \\
 & ={\rm tr}\left[\left(\hat{f}_{1}\hat{f}_{2}\right)^{n-1}\hat{f}_{1}\hat{R}\hat{f}_{2}\right]=-{\rm tr}\left[\left(\hat{\rho}_{k,2}^{{\rm O}\times{\rm USp}}\left(\hat{x},\hat{p};0\right)\right)^{n}\hat{R}\right].
\end{align}
This means that the second condition is also satisfied.

\section{Dualities\label{sec:Dualities}}

In this section we derive exact relations between the matrix models. The formulas which we will use in this section are summarized in Appendix \ref{sec:QM}. Before going on, we provide general features.

The schematic strategy to show the relation is (see also figure \ref{fig:Quivers})
\begin{equation}
\left[{\rm U}\times{\rm U}\right]\rightarrow\hat{A}_{3}=\hat{D}_{3}\leftarrow\left[{\rm O}\times{\rm USp}\right].\label{eq:UU-A3-D3-OSp}
\end{equation}
The equality of the matrix models between the $\hat{A}_{3}$ and $\hat{D}_{3}$ theories is automatically satisfied as \eqref{eq:PF_D3}. On the other hand, the relation between the ${\rm U}\times{\rm U}$ matrix model and the $\hat{A}_{3}$ matrix model is non-trivial. For finding the relation, we apply the Fermi gas formalism for both the matrix models and move to the grand canonical expression. As reviewed in sections \ref{subsec:FGF-UU} and \ref{subsec:FGF-A3}, after applying the Fermi gas formalism, both the ${\rm U}\times{\rm U}$ and $\hat{A}_{3}$ grand partition functions can be written in terms of the Fredholm determinant as \eqref{eq:GPF-UU-FGF}, \eqref{eq:GPF-A3-FGF1} or \eqref{eq:GPF-A3-FGF2}. It is important that the Fredholm determinant ${\rm Det}\left(1+\kappa\hat{{\cal O}}\right)$ is invariant under similarity transformation of $\hat{{\cal O}}$. Hence, if we can show that the density matrices of these theories are equal up to a similarity transformation and an overall constant,
\begin{equation}
\hat{\rho}_{k,M}^{{\rm U}\times{\rm U}}\left(\hat{X},\hat{P};m\right)\sim e^{i\theta}\hat{\rho}_{k',M',F'}^{\hat{A}_{3}}\left(\hat{x},\hat{p};\bm{\zeta}\right),
\end{equation}
we can show that the Fredholm determinants are equal with an appropriate identification of the fugacities, $\kappa\leftrightarrow e^{-i\theta}\kappa$. Here $\sim$ denotes that the both sides are equal up to a similarity transformation. This immediately leads to an exact relation between the ${\rm U}\times{\rm U}$ matrix model and the $\hat{A}_{3}$ matrix model. This is our main strategy in this section. The relation between the ${\rm O}\times{\rm USp}$ matrix model and the $\hat{D}_{3}$ matrix model is also non-trivial, but it can be obtained in the same manner. As reviewed in sections \ref{subsec:FGF-OUSp} and \ref{subsec:FGF-D3}, after applying the Fermi gas formalism, both the ${\rm O}\times{\rm USp}$ and $\hat{D}_{3}$ grand partition functions can be written in terms of the Fredholm determinant as \eqref{eq:GPF-OUSpM0-FGF}, \eqref{eq:GPF-OUSpM1-FGF}, \eqref{eq:GPF-OUSpM2-FGF}, \eqref{eq:GPF-D3-FGF1} or \eqref{eq:GPF-D3-FGF2}. Hence, again, the equality of the density matrix (up to a similarity transformation) immediately leads to the exact relation between the ${\rm O}\times{\rm USp}$ and $\hat{D}_{3}$ matrix models.

As we saw in section \ref{sec:FGF}, the matrix model sometimes has a phase factor which explicitly appears, for example, in \eqref{eq:Z_U-Def}. This is the framing factor, which is caused by a non-trivial framing implicitly chosen in the supersymmetric localization \cite{Witten:1988hf,Kapustin:2009kz} (see also \cite{Marino:2011nm}). We will also see that the matrix model sometimes has a mass-dependent phase factor. Although these phase factors do not play any role for studying dualities (in a sense that the mass-dependent factors of dual theories can be matched by adding appropriate counterterms \cite{Closset:2012vg,Closset:2012vp}), we keep these factors, and hence the final results are exact including these phases.\footnote{The mass-dependent phase plays an important role when, for example, gauging the corresponding flavor symmetry.}

On the other hand, the absolute value of the matrix model is important. Although the Fermi gas formalism provides an $N$-independent overall factor like \eqref{eq:Z_U-Def}, \eqref{eq:Z_O1-Def} or \eqref{eq:Z_O-Def}, which seems to be regarded as a decoupled sector, these factors from both the dual theories non-trivially cancel each other. 

We will check the correspondence of the flavor symmetries by seeing the correspondence of the mass deformation parameters. As commented in section \ref{sec:Matrix-models}, the diagonal subgroup of the ${\rm SU}\left(2\right)\times{\rm SU}\left(2\right)$ flavor symmetry of ${\rm U}\times{\rm U}$ theory corresponds to the ${\rm SU}\left(2\right)$ flavor symmetry of ${\rm O}\times{\rm USp}$ theory \cite{Beratto:2021xmn}. Since we introduced the mass parameters $m_{{\rm UU}}$ and $m_{{\rm OUSp}}$ so that they correspond to the diagonal subgroup of the ${\rm SU}\left(2\right)\times{\rm SU}\left(2\right)$ and the ${\rm SU}\left(2\right)$, respectively, these parameters are expected to satisfy $m_{{\rm UU}}=m_{{\rm OUSp}}$. We will see that this relation indeed holds.

In this section, when we consider the $\left[{\rm U}\times{\rm U}\right]\rightarrow\hat{A}_{3}$ relation, the Chern-Simons level is always $k=4$ for ${\rm U}\times{\rm U}$ and $k=1$ for $\hat{A}_{3}$. Correspondingly, two position and momentum operators, $\left(\hat{X},\hat{P}\right)$ and $\left(\hat{x},\hat{p}\right)$, appear, which commutation relations are $\left[\hat{X},\hat{P}\right]=8\pi i$ and $\left[\hat{x},\hat{p}\right]=2\pi i$. The relation between $\left(\hat{X},\hat{P}\right)$ and $\left(\hat{x},\hat{p}\right)$ depends on the cases. When we consider the $\left[{\rm O}\times{\rm USp}\right]\rightarrow\hat{D}_{3}$ relation, the Chern-Simons level is always $k=1$ for both the ${\rm O}\times{\rm USp}$ and $\hat{D}_{3}$.

\subsection{${\rm U}\left(N\right)_{4}\times{\rm U}\left(N\right)_{-4}\Leftrightarrow{\rm O}\left(2N\right)_{2}\times{\rm USp}\left(2N\right)_{-1}$\label{subsec:DualityM0}}

In this section we study the duality between the ${\rm U}\left(N\right)_{4}\times{\rm U}\left(N\right)_{-4}$ and ${\rm O}\left(2N\right)_{2}\times{\rm USp}\left(2N\right)_{-1}$ theory.
\begin{figure}
\begin{centering}
\includegraphics[scale=0.5]{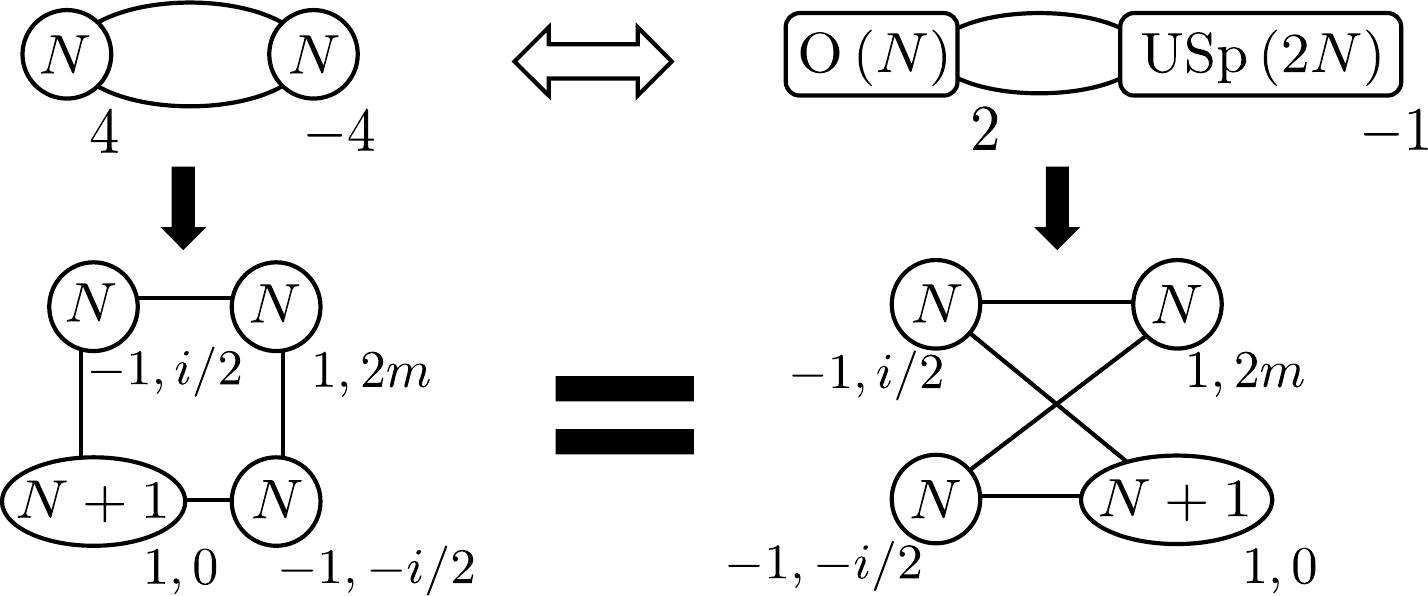}
\par\end{centering}
\caption{Schematic relations between the ${\rm U}\left(N\right)_{4}\times{\rm U}\left(N\right)_{-4}$, $\hat{A}_{3}$, $\hat{D}_{3}$ and ${\rm O}\left(2N\right)_{2}\times{\rm USp}\left(2N\right)_{-1}$ theories. We show an exact relation between the ${\rm U}\left(N\right)_{4}\times{\rm U}\left(N\right)_{-4}$ and ${\rm O}\left(2N\right)_{2}\times{\rm USp}\left(2N\right)_{-1}$ matrix models by embedding them into the $\hat{A}_{3}$ and $\hat{D}_{3}$ matrix models, respectively.\label{fig:DualityM0}}
\end{figure}
Figure \ref{fig:DualityM0} shows \eqref{eq:UU-A3-D3-OSp} in detail.

\subsubsection{From ${\rm U}\left(N\right)_{4}\times{\rm U}\left(N\right)_{-4}$ to $\hat{A}_{3}$}

In this section we show the relation between the ${\rm U}\left(N\right)_{4}\times{\rm U}\left(N\right)_{-4}$ theory and the $\hat{A}_{3}$ theory. As discussed above we start with the density matrix.

The ${\rm U}\times{\rm U}$ density matrix is given by \eqref{eq:DM-UU}. After performing the canonical transformation\footnote{This canonical transformation can be realized by the similarity transformation generated by $e^{\frac{i}{2\hbar}\hat{x}^{2}}e^{\frac{i}{2\hbar}\hat{p}^{2}}e^{\frac{i}{2\hbar}\hat{x}^{2}}$.}
\begin{equation}
\hat{x}\rightarrow-\hat{p},\quad\hat{p}\rightarrow\hat{x},\label{eq:DualM0-Can1}
\end{equation}
it becomes
\begin{equation}
\hat{\rho}_{4,0}^{{\rm U}\times{\rm U}}\left(\hat{X},\hat{P};m\right)\sim\frac{e^{-im\hat{P}}}{2\cosh\frac{\hat{P}}{2}}\frac{e^{im\hat{X}}}{2\cosh\frac{\hat{X}}{2}}.
\end{equation}
This canonical transformation is necessary to match sign of the mass dependent term. To convert this density matrix to the $\hat{A}_{3}$ density matrix, we need four $\cosh^{-1}$ factors. The simplest way to do this is to split the two $\cosh^{-1}$ factors as
\begin{equation}
\hat{\rho}_{4,0}^{{\rm U}\times{\rm U}}\left(2\hat{x},2\hat{p};m\right)\sim\frac{e^{-2im\hat{p}}}{2\cosh\frac{\hat{p}-\frac{1}{2}\pi i}{2}}\frac{1}{2\cosh\frac{\hat{p}+\frac{1}{2}\pi i}{2}}\frac{1}{2\cosh\frac{\hat{x}+\frac{1}{2}\pi i}{2}}\frac{e^{2im\hat{x}}}{2\cosh\frac{\hat{x}-\frac{1}{2}\pi i}{2}},\label{eq:DualityM0-DMUU}
\end{equation}
where
\begin{equation}
\hat{x}=\frac{\hat{X}}{2},\quad\hat{p}=\frac{\hat{P}}{2}.
\end{equation}

On the other hand, the density matrix of the $\hat{A}_{3}$ theory with the parameter (shown in figure \ref{fig:DualityM0})
\begin{equation}
\bm{k}=\left(-1,1,-1,1\right),\quad\bm{N}=\left(N,N,N,N+1\right),\quad\bm{\zeta}=\left(\frac{i}{2},2m,-\frac{i}{2},0\right),\label{eq:DualityM0-para}
\end{equation}
can be read off from \eqref{eq:DM-A31} as
\begin{align}
\hat{\rho}_{1,1,0}^{\hat{A}_{3}}\left(\hat{x},\hat{p};\bm{\zeta}\right) & =e^{-\frac{1}{2}\left(\hat{x}-\hat{p}\right)}\frac{1}{2\cosh\frac{\hat{x}}{2}}e^{2im\left(\hat{x}-\hat{p}\right)}\frac{1}{2\cosh\frac{\hat{p}}{2}}e^{\frac{1}{2}\left(\hat{x}-\hat{p}\right)}\frac{1}{2\cosh\frac{\hat{p}}{2}}\frac{1}{2\cosh\frac{\hat{x}}{2}}.
\end{align}
After an appropriate similarity transformation, we find
\begin{align}
 & \hat{\rho}_{1,1,0}^{\hat{A}_{3}}\left(\hat{x},\hat{p};\bm{\zeta}\right)\nonumber \\
 & \sim e^{-4\pi im^{2}}e^{-2im\hat{p}}e^{-\frac{1}{4}\hat{x}}\frac{1}{2\cosh\frac{\hat{p}}{2}}e^{\frac{1}{2}\hat{x}}\frac{1}{2\cosh\frac{\hat{p}}{2}}e^{-\frac{1}{4}\hat{x}}e^{-\frac{1}{4}\hat{p}}\frac{1}{2\cosh\frac{\hat{x}}{2}}e^{\frac{1}{2}\hat{p}}\frac{1}{2\cosh\frac{\hat{x}}{2}}e^{-\frac{1}{4}\hat{p}}e^{2im\hat{x}}.\label{eq:DualityM0-DMA3}
\end{align}
This is equal to \eqref{eq:DualityM0-DMUU} up to the overall phase. Thus we find the exact relation between the ${\rm U}\left(N\right)_{4}\times{\rm U}\left(N\right)_{-4}$ and $\hat{A}_{3}$ matrix models
\begin{equation}
\hat{\rho}_{4,0}^{{\rm U}\times{\rm U}}\left(2\hat{x},2\hat{p};m\right)\sim e^{4\pi im^{2}}\hat{\rho}_{1,1,0}^{\hat{A}_{3}}\left(\hat{x},\hat{p};\bm{\zeta}\right).\label{eq:DualityM0-DM1}
\end{equation}

Now we move to the grand partition function. The result of the Fermi gas formalism for ${\rm U}\times{\rm U}$ grand partition function is given by \eqref{eq:GPF-UU-FGF} and hence
\begin{equation}
\Xi_{4,0}^{{\rm U}\times{\rm U}}\left(\kappa;m\right)={\rm Det}\left(1+\kappa\hat{\rho}_{4,0}^{{\rm U}\times{\rm U}}\left(2\hat{x},2\hat{p};m\right)\right).
\end{equation}
The result of the Fermi gas formalism for $\hat{A}_{3}$ grand partition function is given by \eqref{eq:GPF-A3-FGF1} in this case and hence
\begin{equation}
\Xi_{1,1,0}^{\hat{A}_{3}}\left(\kappa;\bm{\zeta}\right)={\rm Det}\left(1+\kappa\hat{\rho}_{1,1,0}^{\hat{A}_{3}}\left(\hat{x},\hat{p};\bm{\zeta}\right)\right).
\end{equation}
Note that $Z_{1,1}^{{\rm U}}\left(0\right)=1$. Thus we find an exact relation
\begin{equation}
\Xi_{4,0}^{{\rm U}\times{\rm U}}\left(\kappa;m\right)=\Xi_{1,1,0}^{\hat{A}_{3}}\left(e^{4\pi im^{2}}\kappa;\bm{\zeta}\right).\label{eq:DualityM0-GPF1}
\end{equation}

\subsubsection{From ${\rm O}\left(2N\right)_{2}\times{\rm USp}\left(2N\right)_{-1}$ to $\hat{D}_{3}$}

In this section we show the relation between the ${\rm O}\left(2N\right)_{2}\times{\rm USp}\left(2N\right)_{-1}$ theory and the $\hat{D}_{3}$ theory.

The ${\rm O}\times{\rm USp}$ density matrix is given by \eqref{eq:DM-OUSpM0}
\begin{equation}
\hat{\rho}_{1,0}^{{\rm O}\times{\rm USp}}\left(\hat{x},\hat{p};m\right)=\frac{1}{2\sinh\hat{x}}\hat{R}e^{\frac{i}{4\pi}\hat{x}^{2}}\frac{e^{im\hat{p}}}{2\cosh\frac{\hat{p}}{2}}e^{-\frac{i}{4\pi}\hat{x}^{2}}\hat{R}\left(2\sinh\hat{x}\right)e^{-\frac{i}{4\pi}\hat{x}^{2}}\frac{e^{-im\hat{p}}}{2\cosh\frac{\hat{p}}{2}}e^{\frac{i}{4\pi}\hat{x}^{2}}.
\end{equation}
We can rewrite 
\begin{align}
\hat{\rho}_{1,0}^{{\rm O}\times{\rm USp}}\left(\hat{x},\hat{p};m\right) & \sim\frac{1}{2\sinh\hat{x}}e^{\frac{i}{4\pi}\hat{x}^{2}}\frac{e^{im\hat{p}}}{2\cosh\frac{\hat{p}}{2}}e^{-\frac{i}{4\pi}\hat{x}^{2}}\left(2\sinh\hat{x}\right)e^{-\frac{i}{4\pi}\hat{x}^{2}}\frac{e^{im\hat{p}}}{2\cosh\frac{\hat{p}}{2}}e^{\frac{i}{4\pi}\hat{x}^{2}}\nonumber \\
 & =e^{2i\pi m^{2}}\frac{1}{2\sinh\hat{x}}e^{\frac{i}{4\pi}\hat{x}^{2}}\frac{1}{2\cosh\frac{\hat{p}}{2}}e^{-\frac{i}{4\pi}\hat{x}^{2}}e^{im\hat{p}}\left(2\sinh\hat{x}\right)e^{im\hat{p}}e^{-\frac{i}{4\pi}\hat{x}^{2}}\frac{1}{2\cosh\frac{\hat{p}}{2}}e^{\frac{i}{4\pi}\hat{x}^{2}}.
\end{align}
By performing the canonical transformation
\begin{equation}
\hat{x}\rightarrow-\hat{p},\quad\hat{p}\rightarrow\hat{x},\label{eq:DualM0-Can2}
\end{equation}
we obtain
\begin{align}
\hat{\rho}_{1,0}^{{\rm O}\times{\rm USp}}\left(\hat{x},\hat{p};m\right) & \sim e^{2i\pi m^{2}}\frac{1}{2\sinh\hat{p}}e^{\frac{i}{4\pi}\hat{p}^{2}}\frac{1}{2\cosh\frac{\hat{x}}{2}}e^{-\frac{i}{4\pi}\hat{p}^{2}}e^{im\hat{x}}\left(2\sinh\hat{p}\right)e^{im\hat{x}}e^{-\frac{i}{4\pi}\hat{p}^{2}}\frac{1}{2\cosh\frac{\hat{x}}{2}}e^{\frac{i}{4\pi}\hat{p}^{2}}\nonumber \\
 & \sim e^{2i\pi m^{2}}\frac{1}{2\sinh\hat{p}}e^{-\frac{i}{4\pi}\hat{x}^{2}}\frac{1}{2\cosh\frac{\hat{p}}{2}}e^{\frac{i}{4\pi}\hat{x}^{2}}e^{im\hat{x}}\left(2\sinh\hat{p}\right)e^{im\hat{x}}e^{\frac{i}{4\pi}\hat{x}^{2}}\frac{1}{2\cosh\frac{\hat{p}}{2}}e^{-\frac{i}{4\pi}\hat{x}^{2}}.
\end{align}
Here we used the equality
\begin{equation}
e^{\pm\frac{i}{4\pi}\hat{p}^{2}}\frac{1}{2\cosh\frac{\hat{x}}{2}}e^{\mp\frac{i}{4\pi}\hat{p}^{2}}=e^{\mp\frac{i}{4\pi}\hat{x}^{2}}\frac{1}{2\cosh\frac{\hat{p}}{2}}e^{\pm\frac{i}{4\pi}\hat{x}^{2}}.\label{eq:Sim1}
\end{equation}
This equality comes from the second line of \eqref{eq:OpSim}.

On the other hand, the $\hat{D}_{3}$ density matrix with the parameter \eqref{eq:DualityM0-para} can be read off from \eqref{eq:DM-D31} as
\begin{align}
\hat{\rho}_{1,1,0}^{\hat{D}_{3}}\left(\hat{x},\hat{p};\bm{\zeta}\right) & =\frac{i}{2}e^{-\frac{i}{4\pi}\hat{p}^{2}}e^{-\frac{i}{4\pi}\hat{x}^{2}}\frac{1}{\tanh\frac{\hat{p}}{2}}e^{-\frac{i}{4\pi}\hat{x}^{2}}e^{-\frac{i}{4\pi}\hat{p}^{2}}\frac{1}{2\cosh\frac{\hat{x}}{2}}\frac{1}{2\cosh\frac{\hat{p}}{2}}2\sinh\frac{\hat{x}}{2}\nonumber \\
 & \quad\times e^{\frac{i}{4\pi}\hat{p}^{2}}e^{\frac{i}{4\pi}\hat{x}^{2}}\left\{ \frac{1}{\tanh\frac{\hat{p}}{2}},e^{2im\hat{x}}\right\} e^{\frac{i}{4\pi}\hat{x}^{2}}e^{\frac{i}{4\pi}\hat{p}^{2}}2\sinh\frac{\hat{x}}{2}\frac{1}{2\cosh\frac{\hat{p}}{2}}\frac{1}{2\cosh\frac{\hat{x}}{2}}.
\end{align}
We can rewrite
\begin{align}
\hat{\rho}_{1,1,0}^{\hat{D}_{3}}\left(\hat{x},\hat{p};\bm{\zeta}\right) & \sim-\frac{1}{2}e^{-\frac{i}{4\pi}\hat{p}^{2}}e^{-\frac{i}{4\pi}\hat{x}^{2}}\frac{1}{2\sinh\hat{p}}e^{-\frac{i}{4\pi}\hat{x}^{2}}e^{-\frac{i}{4\pi}\hat{p}^{2}}\frac{1}{2\cosh\frac{\hat{p}}{2}}\nonumber \\
 & \quad\times e^{\frac{i}{4\pi}\hat{p}^{2}}e^{\frac{i}{4\pi}\hat{x}^{2}}2\sinh\frac{\hat{p}}{2}\left\{ \frac{1}{\tanh\frac{\hat{p}}{2}},e^{2im\hat{x}}\right\} 2\sinh\frac{\hat{p}}{2}e^{\frac{i}{4\pi}\hat{x}^{2}}e^{\frac{i}{4\pi}\hat{p}^{2}}\frac{1}{2\cosh\frac{\hat{p}}{2}}.
\end{align}
One can show that
\begin{equation}
2\sinh\frac{\hat{p}}{2}\left\{ \frac{1}{\tanh\frac{\hat{p}}{2}},e^{2im\hat{x}}\right\} 2\sinh\frac{\hat{p}}{2}=2e^{im\hat{x}}\left(2\sinh\hat{p}\right)e^{im\hat{x}}.
\end{equation}
By using this relation, we obtain
\begin{align}
\hat{\rho}_{1,1,0}^{\hat{D}_{3}}\left(\hat{x},\hat{p};\bm{\zeta}\right) & \sim\frac{1}{2\sinh\hat{p}}e^{-\frac{i}{4\pi}\hat{x}^{2}}\frac{1}{2\cosh\frac{\hat{p}}{2}}e^{\frac{i}{4\pi}\hat{x}^{2}}e^{im\hat{x}}\left(2\sinh\hat{p}\right)e^{im\hat{x}}e^{\frac{i}{4\pi}\hat{x}^{2}}\frac{1}{2\cosh\frac{\hat{p}}{2}}e^{-\frac{i}{4\pi}\hat{x}^{2}}.
\end{align}
Thus we find the exact relation between the ${\rm O}\left(2N\right)_{2}\times{\rm USp}\left(2N\right)_{-1}$ and $\hat{D}_{3}$ matrix models
\begin{equation}
\hat{\rho}_{1,0}^{{\rm O}\times{\rm USp}}\left(\hat{x},\hat{p};m\right)\sim e^{2i\pi m^{2}}\hat{\rho}_{1,1,0}^{\hat{D}_{3}}\left(\hat{x},\hat{p};\bm{\zeta}\right).\label{eq:DualityM0-DM2}
\end{equation}

Now we move to the grand partition function. The result of the Fermi gas formalism for ${\rm O}\times{\rm USp}$ grand partition function is given by \eqref{eq:GPF-OUSpM0-FGF} in this case and hence
\begin{equation}
\Xi_{1,0}^{{\rm O}\times{\rm USp}}\left(\kappa;m\right)=\sqrt{{\rm Det}\left(1+\kappa\hat{\rho}_{1,0}^{{\rm O}\times{\rm USp}}\left(\hat{x},\hat{p};m\right)\right)}.
\end{equation}
The result of the Fermi gas formalism for $\hat{D}_{3}$ grand partition function is given by \eqref{eq:GPF-D3-FGF1} in this case and hence
\begin{equation}
\Xi_{1,1,0}^{\hat{D}_{3}}\left(\kappa;\bm{\zeta}\right)=\sqrt{{\rm Det}\left(1+\kappa\hat{\rho}_{1,1,0}^{\hat{D}_{3}}\left(\hat{x},\hat{p};\bm{\zeta}\right)\right)}.
\end{equation}
Note that $Z_{1,1}^{{\rm U}}\left(0\right)=1$. Thus we find an exact relation
\begin{equation}
\Xi_{1,0}^{{\rm O}\times{\rm USp}}\left(\kappa;m\right)=\Xi_{1,1,0}^{\hat{D}_{3}}\left(e^{2i\pi m^{2}}\kappa;\bm{\zeta}\right).\label{eq:DualityM0-GPF2}
\end{equation}

By combining this result and \eqref{eq:DualityM0-GPF1}, we obtain
\begin{equation}
\Xi_{4,0}^{{\rm U}\times{\rm U}}\left(\kappa;m\right)=\Xi_{1,0}^{{\rm O}\times{\rm USp}}\left(e^{2i\pi m^{2}}\kappa;m\right).
\end{equation}
This equality immediately leads to the final result
\begin{equation}
Z_{4,\left(N,N\right)}^{{\rm U}\times{\rm U}}\left(m\right)=e^{2i\pi m^{2}N}Z_{k,\left(2N,2N\right)}^{{\rm O}\times{\rm USp}}\left(m\right).\label{eq:DualityM0-PF}
\end{equation}
Therefore, we have rigorously checked the ${\rm U}\left(N\right)_{4}\times{\rm U}\left(N\right)_{-4}\Leftrightarrow{\rm O}\left(2N\right)_{2}\times{\rm USp}\left(2N\right)_{-1}$ duality by proving the exact equality of the absolute values of the matrix models for arbitrary rank $N$. We have also shown that the correspondence of the mass parameters is the expected one.

\subsection{${\rm U}\left(N+1\right)_{4}\times{\rm U}\left(N\right)_{-4}\Leftrightarrow{\rm O}\left(2N+1\right)_{2}\times{\rm USp}\left(2N\right)_{-1}$\label{subsec:DualityM1}}

In this section we study the duality between the ${\rm U}\left(N+1\right)_{4}\times{\rm U}\left(N\right)_{-4}$ and ${\rm O}\left(2N+1\right)_{2}\times{\rm USp}\left(2N\right)_{-1}$ theory.
\begin{figure}
\begin{centering}
\includegraphics[scale=0.5]{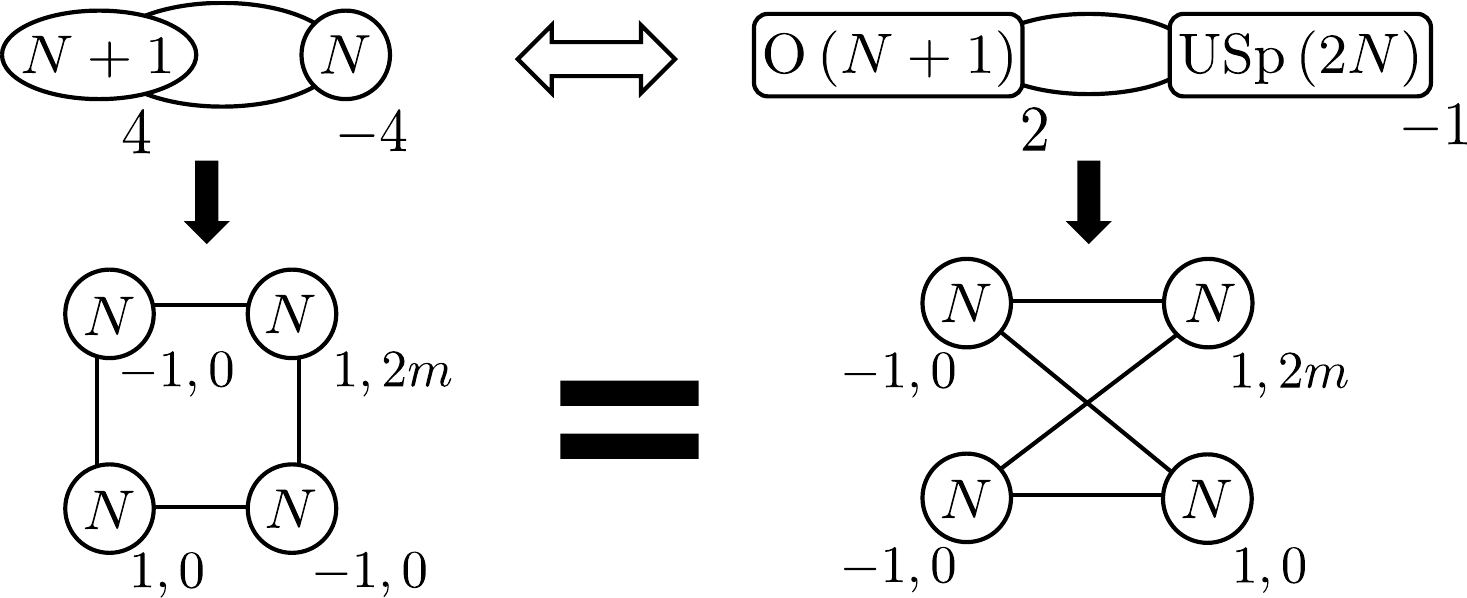}
\par\end{centering}
\caption{Schematic relations between the ${\rm U}\left(N+1\right)_{4}\times{\rm U}\left(N\right)_{-4}$, $\hat{A}_{3}$, $\hat{D}_{3}$ and ${\rm O}\left(2N+1\right)_{2}\times{\rm USp}\left(2N\right)_{-1}$ theories. We show an exact relation between the ${\rm U}\left(N+1\right)_{4}\times{\rm U}\left(N\right)_{-4}$ and ${\rm O}\left(2N+1\right)_{2}\times{\rm USp}\left(2N\right)_{-1}$ matrix models by embedding then into the $\hat{A}_{3}$ and $\hat{D}_{3}$ matrix models, respectively.\label{fig:DualityM1}}
\end{figure}
Figure \ref{fig:DualityM1} shows \eqref{eq:UU-A3-D3-OSp} in detail.

\subsubsection{From ${\rm U}\left(N+1\right)_{4}\times{\rm U}\left(N\right)_{-4}$ to $\hat{A}_{3}$}

In this section we show the relation between the ${\rm U}\left(N+1\right)_{4}\times{\rm U}\left(N\right)_{-4}$ theory and the $\hat{A}_{3}$ theory.

The ${\rm U}\times{\rm U}$ density matrix can be read off from \eqref{eq:DM-UU} as
\begin{equation}
\hat{\rho}_{4,1}^{{\rm U}\times{\rm U}}\left(\hat{X},\hat{P};m\right)=e^{im\hat{X}}\frac{2\sinh\frac{\hat{X}}{8}}{2\sinh\frac{\hat{X}}{2}}\frac{1}{2\cosh\frac{\hat{P}}{2}}\frac{1}{2\cosh\frac{\hat{X}}{8}}e^{im\hat{P}}.
\end{equation}
For relating this density matrix to the $\hat{A}_{3}$ density matrix, we need to get rid of the hyperbolic sine function at the numerator. This factor can be canceled by using $\sinh2y=2\sinh y\cosh y$ twice to the denominator. Then, by using the relation
\begin{equation}
\frac{1}{2\cosh\frac{\hat{X}}{8}}\frac{1}{2\cosh\frac{\hat{P}}{2}}\frac{1}{2\cosh\frac{\hat{X}}{8}}=\frac{1}{2\cosh\frac{\hat{P}}{4}}\frac{1}{2\cosh\frac{\hat{X}}{4}}\frac{1}{2\cosh\frac{\hat{P}}{4}},
\end{equation}
we find
\begin{equation}
\hat{\rho}_{4,1}^{{\rm U}\times{\rm U}}\left(2\hat{x},2\hat{p};m\right)=\frac{e^{2im\hat{x}}}{2\cosh\frac{\hat{x}}{2}}\frac{1}{2\cosh\frac{\hat{p}}{2}}\frac{1}{2\cosh\frac{\hat{x}}{2}}\frac{e^{2im\hat{p}}}{2\cosh\frac{\hat{p}}{2}},
\end{equation}
where
\begin{equation}
\hat{x}=\frac{\hat{X}}{2},\quad\hat{p}=\frac{\hat{P}}{2}.
\end{equation}
By performing the canonical transformation as well as the ${\rm U}\left(N\right)_{4}\times{\rm U}\left(N\right)_{-4}$ case
\begin{equation}
\hat{x}\rightarrow-\hat{p},\quad\hat{p}\rightarrow\hat{x},
\end{equation}
we obtain
\begin{equation}
\hat{\rho}_{4,1}^{{\rm U}\times{\rm U}}\left(2\hat{x},2\hat{p};m\right)=\frac{e^{-2im\hat{p}}}{2\cosh\frac{\hat{p}}{2}}\frac{1}{2\cosh\frac{\hat{x}}{2}}\frac{1}{2\cosh\frac{\hat{p}}{2}}\frac{e^{2im\hat{x}}}{2\cosh\frac{\hat{x}}{2}}.
\end{equation}
This canonical transformation is necessary to match sign of the mass dependent term.

On the other hand, the density matrix of the $\hat{A}_{3}$ theory with the parameter shown in figure \ref{fig:DualityM1}
\begin{equation}
\bm{k}=\left(-1,1,-1,1\right),\quad\bm{N}=\left(N,N,N,N\right),\quad\bm{\zeta}=\left(0,2m,0,0\right),\label{eq:DualityM1-para}
\end{equation}
can be read off from \eqref{eq:DM-A31} as
\begin{align}
\hat{\rho}_{1,0,0}^{\hat{A}_{3}}\left(\hat{x},\hat{p};\bm{\zeta}\right) & =\frac{1}{2\cosh\frac{\hat{x}}{2}}e^{2im\left(\hat{x}-\hat{p}\right)}\frac{1}{2\cosh\frac{\hat{p}}{2}}\frac{1}{2\cosh\frac{\hat{x}}{2}}\frac{1}{2\cosh\frac{\hat{p}}{2}}.
\end{align}
After an appropriate similarity transformation, we find
\begin{equation}
\hat{\rho}_{1,0,0}^{\hat{A}_{3}}\left(\hat{x},\hat{p};\bm{\zeta}\right)\sim e^{-4\pi im^{2}}\frac{e^{-2im\hat{p}}}{2\cosh\frac{\hat{p}}{2}}\frac{1}{2\cosh\frac{\hat{x}}{2}}\frac{1}{2\cosh\frac{\hat{p}}{2}}\frac{e^{2im\hat{x}}}{2\cosh\frac{\hat{x}}{2}}.
\end{equation}
Thus we find the exact relation between the ${\rm U}\left(N+1\right)_{4}\times{\rm U}\left(N\right)_{-4}$ and $\hat{A}_{3}$ matrix models
\begin{equation}
\hat{\rho}_{4,1}^{{\rm U}\times{\rm U}}\left(2\hat{x},2\hat{p};m\right)\sim e^{4\pi im^{2}}\hat{\rho}_{1,0,0}^{\hat{A}_{3}}\left(\hat{x},\hat{p};\bm{\zeta}\right).\label{eq:DualityM1-DM1}
\end{equation}

Now we move to the grand partition function. The result of the Fermi gas formalism for ${\rm U}\times{\rm U}$ grand partition function is given by \eqref{eq:GPF-UU-FGF} and hence
\begin{equation}
\Xi_{4,1}^{{\rm U}\times{\rm U}}\left(\kappa;m\right)=Z_{4,1}^{{\rm U}}\left(0\right){\rm Det}\left(1+\kappa\hat{\rho}_{4,1}^{{\rm U}\times{\rm U}}\left(2\hat{x},2\hat{p};m\right)\right).
\end{equation}
The result of the Fermi gas formalism for $\hat{A}_{3}$ grand partition function is given by \eqref{eq:GPF-A3-FGF1} in this case and hence
\begin{equation}
\Xi_{1,0,0}^{\hat{A}_{3}}\left(\kappa;\bm{\zeta}\right)={\rm Det}\left(1+\kappa\hat{\rho}_{1,0,0}^{\hat{A}_{3}}\left(\hat{x},\hat{p};\bm{\zeta}\right)\right).
\end{equation}
Thus we find an exact relation
\begin{equation}
\frac{\Xi_{4,1}^{{\rm U}\times{\rm U}}\left(\kappa;m\right)}{Z_{4,1}^{{\rm U}}\left(0\right)}=\Xi_{1,0,0}^{\hat{A}_{3}}\left(e^{4\pi im^{2}}\kappa;\bm{\zeta}\right).\label{eq:DualityM1-GPF1}
\end{equation}

\subsubsection{From ${\rm O}\left(2N+1\right)_{2}\times{\rm USp}\left(2N\right)_{-1}$ to $\hat{D}_{3}$}

In this section we show the relation between the ${\rm O}\left(2N+1\right)_{2}\times{\rm USp}\left(2N\right)_{-1}$ theory and the $\hat{D}_{3}$ theory.

The ${\rm O}\times{\rm USp}$ density matrix is given by \eqref{eq:DM-OUSpM1}
\begin{align}
\hat{\rho}_{1,1}^{{\rm O}\times{\rm USp}}\left(\hat{x},\hat{p};m\right) & =\tanh\frac{\hat{x}}{2}\hat{R}e^{\frac{i}{4\pi}\hat{x}^{2}}\frac{e^{im\hat{p}}}{2\cosh\frac{\hat{p}}{2}}e^{-\frac{i}{4\pi}\hat{x}^{2}}\nonumber \\
 & \quad\times\hat{R}\frac{2\sinh\hat{x}}{\left(2\cosh\frac{\hat{x}+2\pi m}{2}\right)\left(2\cosh\frac{\hat{x}-2\pi m}{2}\right)}e^{-\frac{i}{4\pi}\hat{x}^{2}}\frac{e^{-im\hat{p}}}{2\cosh\frac{\hat{p}}{2}}e^{\frac{i}{4\pi}\hat{x}^{2}}.
\end{align}
After performing the canonical transformation as well as the ${\rm O}\left(2N\right)_{2}\times{\rm USp}\left(2N\right)_{-1}$ case
\begin{equation}
\hat{x}\rightarrow-\hat{p},\quad\hat{p}\rightarrow\hat{x},
\end{equation}
we obtain
\begin{align}
\hat{\rho}_{1,0}^{{\rm O}\times{\rm USp}}\left(\hat{x},\hat{p};m\right) & \sim e^{2i\pi m^{2}}\tanh\frac{\hat{p}}{2}e^{-\frac{i}{4\pi}\hat{x}^{2}}\frac{1}{2\cosh\frac{\hat{p}}{2}}e^{\frac{i}{4\pi}\hat{x}^{2}}\nonumber \\
 & \quad\times e^{im\hat{x}}\frac{2\sinh\hat{p}}{\left(2\cosh\frac{\hat{p}+2\pi m}{2}\right)\left(2\cosh\frac{\hat{p}-2\pi m}{2}\right)}e^{im\hat{x}}e^{\frac{i}{4\pi}\hat{x}^{2}}\frac{1}{2\cosh\frac{\hat{p}}{2}}e^{-\frac{i}{4\pi}\hat{x}^{2}}.
\end{align}
Here we used \eqref{eq:Sim1}.

On the other hand, the $\hat{D}_{3}$ density matrix with the parameter \eqref{eq:DualityM1-para} can be read off from \eqref{eq:DM-D31} as
\begin{align}
\hat{\rho}_{1,0,0}^{\hat{D}_{3}}\left(\hat{x},\hat{p};\bm{\zeta}\right) & =\frac{1}{2}e^{-\frac{i}{4\pi}\hat{p}^{2}}e^{-\frac{i}{4\pi}\hat{x}^{2}}\tanh\frac{\hat{p}}{2}e^{-\frac{i}{4\pi}\hat{x}^{2}}e^{-\frac{i}{4\pi}\hat{p}^{2}}\frac{1}{2\cosh\frac{\hat{p}}{2}}\nonumber \\
 & \quad\times e^{\frac{i}{4\pi}\hat{p}^{2}}e^{\frac{i}{4\pi}\hat{x}^{2}}\left\{ \tanh\frac{\hat{p}}{2},e^{2im\hat{x}}\right\} e^{\frac{i}{4\pi}\hat{x}^{2}}e^{\frac{i}{4\pi}\hat{p}^{2}}\frac{1}{2\cosh\frac{\hat{p}}{2}}.
\end{align}
We can rewrite
\begin{equation}
\hat{\rho}_{1,0,0}^{\hat{D}_{3}}\left(\hat{x},\hat{p};\bm{\zeta}\right)\sim\frac{1}{2}e^{-\frac{i}{4\pi}\hat{x}^{2}}\tanh\frac{\hat{p}}{2}e^{-\frac{i}{4\pi}\hat{x}^{2}}\frac{1}{2\cosh\frac{\hat{p}}{2}}e^{\frac{i}{4\pi}\hat{x}^{2}}\left\{ \tanh\frac{\hat{p}}{2},e^{2im\hat{x}}\right\} e^{\frac{i}{4\pi}\hat{x}^{2}}\frac{1}{2\cosh\frac{\hat{p}}{2}}.
\end{equation}
One can show that
\begin{equation}
\frac{1}{2}\left\{ \tanh\frac{\hat{p}}{2},e^{2im\hat{x}}\right\} =e^{im\hat{x}}\frac{2\sinh\hat{p}}{\left(2\cosh\frac{\hat{p}+2\pi m}{2}\right)\left(2\cosh\frac{\hat{p}-2\pi m}{2}\right)}e^{im\hat{x}}.
\end{equation}
By using this relation, we obtain
\begin{align}
 & \hat{\rho}_{1,0,0}^{\hat{D}_{3}}\left(\hat{x},\hat{p};\bm{\zeta}\right)\nonumber \\
 & \sim e^{-\frac{i}{4\pi}\hat{x}^{2}}\tanh\frac{\hat{p}}{2}e^{-\frac{i}{4\pi}\hat{x}^{2}}\frac{1}{2\cosh\frac{\hat{p}}{2}}e^{\frac{i}{4\pi}\hat{x}^{2}}e^{im\hat{x}}\frac{2\sinh\hat{p}}{\left(2\cosh\frac{\hat{p}+2\pi m}{2}\right)\left(2\cosh\frac{\hat{p}-2\pi m}{2}\right)}e^{im\hat{x}}e^{\frac{i}{4\pi}\hat{x}^{2}}\frac{1}{2\cosh\frac{\hat{p}}{2}}.
\end{align}
Thus we find the exact relation between the ${\rm O}\left(2N+1\right)_{2}\times{\rm USp}\left(2N\right)_{-1}$ and $\hat{D}_{3}$ matrix models
\begin{equation}
\hat{\rho}_{1,1}^{{\rm O}\times{\rm USp}}\left(\hat{x},\hat{p};m\right)\sim e^{2i\pi m^{2}}\hat{\rho}_{1,0,0}^{\hat{D}_{3}}\left(\hat{x},\hat{p};\bm{\zeta}\right).\label{eq:DualityM1-DM2}
\end{equation}

The result of the Fermi gas formalism for ${\rm O}\times{\rm USp}$ grand partition function is given by \eqref{eq:GPF-OUSpM1-FGF} in this case and hence
\begin{equation}
\Xi_{1,1}^{{\rm O}\times{\rm USp}}\left(\kappa;m\right)=Z_{1}^{{\rm O}}\sqrt{{\rm Det}\left(1+\kappa\hat{\rho}_{1,1}^{{\rm O}\times{\rm USp}}\left(\hat{x},\hat{p};m\right)\right)}.
\end{equation}
The result of the Fermi gas formalism for $\hat{D}_{3}$ grand partition function is given by \eqref{eq:GPF-D3-FGF1} in this case and hence
\begin{equation}
\Xi_{1,0,0}^{\hat{D}_{3}}\left(\kappa;\bm{\zeta}\right)=\sqrt{{\rm Det}\left(1+\kappa\hat{\rho}_{1,0,0}^{\hat{D}_{3}}\left(\hat{x},\hat{p};\bm{\zeta}\right)\right)}.
\end{equation}
Thus we find an exact relation
\begin{equation}
\frac{\Xi_{1,1}^{{\rm O}\times{\rm USp}}\left(\kappa;m\right)}{Z_{1}^{{\rm O}}}=\Xi_{1,0,0}^{\hat{D}_{3}}\left(e^{2i\pi m^{2}}\kappa;\bm{\zeta}\right).\label{eq:DualityM1-GPF2}
\end{equation}

By combining this result and \eqref{eq:DualityM1-GPF1}, we obtain
\begin{equation}
\frac{\Xi_{4,1}^{{\rm U}\times{\rm U}}\left(\kappa;m\right)}{Z_{4,1}^{{\rm U}}\left(0\right)}=\frac{\Xi_{1,0}^{{\rm O}\times{\rm USp}}\left(e^{2i\pi m^{2}}\kappa;m\right)}{Z_{1}^{{\rm O}}}.
\end{equation}
This equality immediately leads to the final result
\begin{equation}
\frac{Z_{4,\left(N+1,N\right)}^{{\rm U}\times{\rm U}}\left(m\right)}{Z_{4,1}^{{\rm U}}\left(0\right)}=e^{2i\pi m^{2}N}\frac{Z_{1,\left(2N+1,2N\right)}^{{\rm O}\times{\rm USp}}\left(m\right)}{Z_{1}^{{\rm O}}}.\label{eq:DualityM1-PF}
\end{equation}
The correspondence of the mass parameters is the expected one as well as the case of the ${\rm U}\left(N\right)_{4}\times{\rm U}\left(N\right)_{-4}\Leftrightarrow{\rm O}\left(2N\right)_{2}\times{\rm USp}\left(2N\right)_{-1}$ duality.

On the other hand, at first sight the relation \eqref{eq:DualityM1-PF} would imply that an additional ${\rm U}\left(1\right)_{4}$ decoupled topological sector (which corresponds to $Z_{4,1}^{{\rm U}}\left(0\right)$) and an additional ${\rm O}\left(1\right)$ decoupled topological sector (which corresponds to $Z_{1}^{{\rm O}}$) are necessary in the right and left side, respectively. However, interestingly, the values of both of the factors are equal since 
\begin{equation}
Z_{4,1}^{{\rm U}}\left(0\right)=Z_{1}^{{\rm O}}=\frac{1}{2}.
\end{equation}
Hence the neither decoupled sector is necessary at least from the viewpoint of the partition function.\footnote{The study of the index also suggests that there is no decoupled sector \cite{Beratto:2021xmn}.} Therefore, we have rigorously checked the ${\rm U}\left(N+1\right)_{4}\times{\rm U}\left(N\right)_{-4}\Leftrightarrow{\rm O}\left(2N+1\right)_{2}\times{\rm USp}\left(2N\right)_{-1}$ duality by proving the exact equality of the absolute values of the matrix models for arbitrary rank $N$.

\subsection{${\rm U}\left(N+2\right)_{4}\times{\rm U}\left(N\right)_{-4}\Leftrightarrow{\rm O}\left(2N+2\right)_{2}\times{\rm USp}\left(2N\right)_{-1}$\label{subsec:DualityM2}}

In this section we study the duality between the ${\rm U}\left(N+2\right)_{4}\times{\rm U}\left(N\right)_{-4}$ and ${\rm O}\left(2N+2\right)_{2}\times{\rm USp}\left(2N\right)_{-1}$ theory. For this duality we do not consider the mass deformation, namely we set $m=0$.
\begin{figure}
\begin{centering}
\includegraphics[scale=0.5]{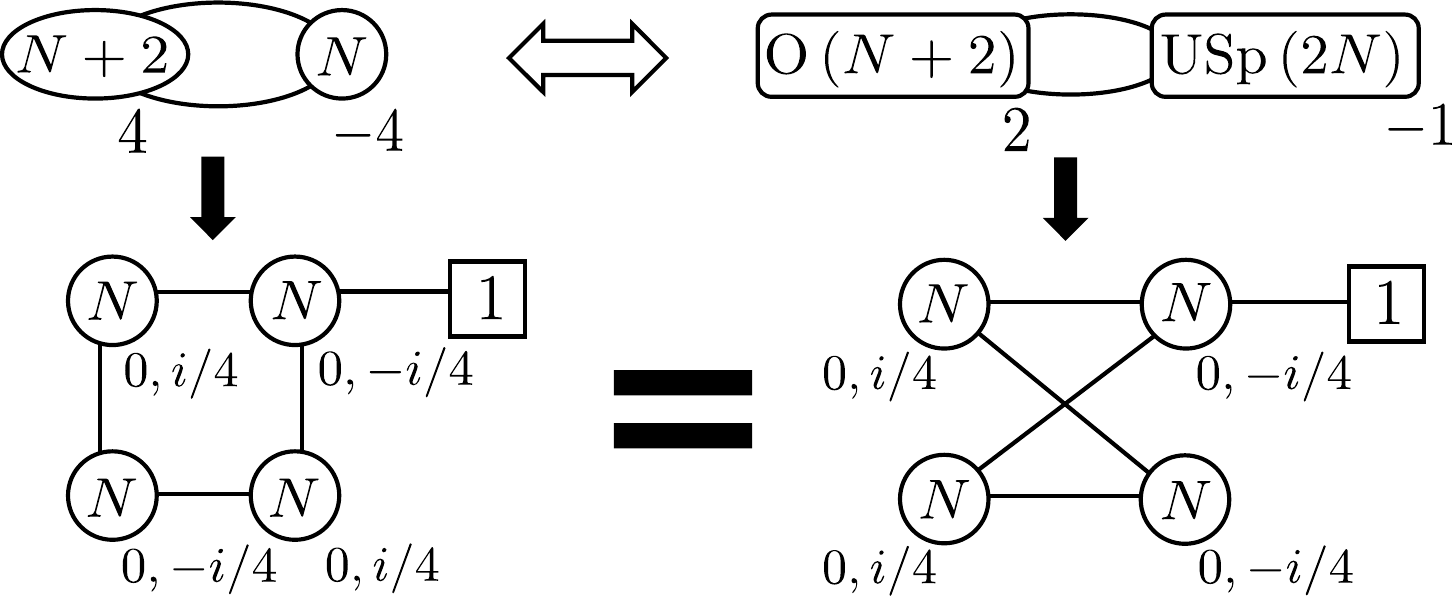}
\par\end{centering}
\caption{Schematic relations between the ${\rm U}\left(N+2\right)_{4}\times{\rm U}\left(N\right)_{-4}$, $\hat{A}_{3}$, $\hat{D}_{3}$ and ${\rm O}\left(2N+2\right)_{2}\times{\rm USp}\left(2N\right)_{-1}$ theories. We show an exact relation between the ${\rm U}\left(N+2\right)_{4}\times{\rm U}\left(N\right)_{-4}$ and ${\rm O}\left(2N+2\right)_{2}\times{\rm USp}\left(2N\right)_{-1}$ matrix models by embedding then into the $\hat{A}_{3}$ and $\hat{D}_{3}$ matrix models, respectively.\label{fig:DualityM2}}
\end{figure}
Figure \ref{fig:DualityM2} shows \eqref{eq:UU-A3-D3-OSp} in detail.

\subsubsection{From ${\rm U}\left(N+2\right)_{4}\times{\rm U}\left(N\right)_{-4}$ to $\hat{A}_{3}$}

In this section we show the relation between the ${\rm U}\left(N+2\right)_{4}\times{\rm U}\left(N\right)_{-4}$ theory and the $\hat{A}_{3}$ theory.

The ${\rm U}\times{\rm U}$ density matrix is given by \eqref{eq:DM-UU}
\begin{equation}
\hat{\rho}_{4,2}^{{\rm U}\times{\rm U}}\left(\hat{X},\hat{P};0\right)=\frac{1}{2\cosh\frac{\hat{X}+i\pi}{8}}\frac{1}{2\cosh\frac{\hat{X}-i\pi}{8}}\frac{1}{2\cosh\frac{\hat{p}}{2}}\frac{1}{2\cosh\frac{\hat{X}+i\pi}{8}}\frac{1}{2\cosh\frac{\hat{X}-i\pi}{8}}.
\end{equation}
Because in this case we have four $\cosh\frac{\hat{X}}{8}$ factors, we just rescale $\hat{X}$ as
\begin{equation}
\hat{\rho}_{4,2}^{{\rm U}\times{\rm U}}\left(4\hat{x},\hat{p};0\right)=\frac{1}{2\cosh\frac{\hat{x}+\frac{1}{4}i\pi}{2}}\frac{1}{2\cosh\frac{\hat{x}-\frac{1}{4}i\pi}{2}}\frac{1}{2\cosh\frac{\hat{p}}{2}}\frac{1}{2\cosh\frac{\hat{x}+\frac{1}{4}i\pi}{2}}\frac{1}{2\cosh\frac{\hat{x}-\frac{1}{4}i\pi}{2}},
\end{equation}
where
\begin{equation}
\hat{x}=\frac{\hat{X}}{4},\quad\hat{p}=\hat{P}.
\end{equation}
By performing the canonical transformation as well as the ${\rm U}\left(N\right)_{4}\times{\rm U}\left(N\right)_{-4}$ case
\begin{equation}
\hat{x}\rightarrow-\hat{p},\quad\hat{p}\rightarrow\hat{x},
\end{equation}
we obtain
\begin{equation}
\hat{\rho}_{4,2}^{{\rm U}\times{\rm U}}\left(4\hat{x},\hat{p};0\right)=\frac{1}{2\cosh\frac{\hat{p}-\frac{1}{4}i\pi}{2}}\frac{1}{2\cosh\frac{\hat{p}+\frac{1}{4}i\pi}{2}}\frac{1}{2\cosh\frac{\hat{x}}{2}}\frac{1}{2\cosh\frac{\hat{p}-\frac{1}{4}i\pi}{2}}\frac{1}{2\cosh\frac{\hat{p}+\frac{1}{4}i\pi}{2}}.
\end{equation}

On the other hand, the density matrix of the $\hat{A}_{3}$ theory with the parameter shown in figure \ref{fig:DualityM2}
\begin{equation}
\bm{k}=\left(0,0,0,0\right),\quad\bm{N}=\left(N,N,N,N\right),\quad\bm{\zeta}=\left(\frac{i}{4},-\frac{i}{4},\frac{i}{4},-\frac{i}{4}\right),\label{eq:DualityM2-para}
\end{equation}
can be read off from \eqref{eq:DM-A32} as
\begin{equation}
\hat{\rho}_{0,0,1}^{\hat{A}_{3}}\left(\hat{x},\hat{p};\bm{\zeta}\right)=e^{-\frac{\pi}{2\hbar}\hat{x}}\frac{1}{2\cosh\frac{\hat{p}}{2}}e^{\frac{\pi}{2\hbar}\hat{x}}\frac{1}{2\cosh\frac{\hat{x}}{2}}\frac{1}{2\cosh\frac{\hat{p}}{2}}e^{-\frac{\pi}{2\hbar}\hat{x}}\frac{1}{2\cosh\frac{\hat{p}}{2}}e^{\frac{\pi}{2\hbar}\hat{x}}\frac{1}{2\cosh\frac{\hat{p}}{2}}.
\end{equation}
Thus we find the exact relation between the ${\rm U}\left(N+2\right)_{4}\times{\rm U}\left(N\right)_{-4}$ and $\hat{A}_{3}$ matrix models
\begin{equation}
\hat{\rho}_{4,2}^{{\rm U}\times{\rm U}}\left(4\hat{x},\hat{p};0\right)\sim\hat{\rho}_{0,0,1}^{\hat{A}_{3}}\left(\hat{x},\hat{p};\bm{\zeta}\right).\label{eq:DualityM2-DM1}
\end{equation}

Now we move to the grand partition function. The result of the Fermi gas formalism for ${\rm U}\times{\rm U}$ grand partition function is given by \eqref{eq:GPF-UU-FGF} and hence
\begin{equation}
\Xi_{4,2}^{{\rm U}\times{\rm U}}\left(\kappa;0\right)=Z_{4,2}^{{\rm U}}\left(0\right){\rm Det}\left(1+\kappa\hat{\rho}_{4,2}^{{\rm U}\times{\rm U}}\left(4\hat{x},\hat{p};0\right)\right).
\end{equation}
The result of the Fermi gas formalism for $\hat{A}_{3}$ grand partition function is given by \eqref{eq:GPF-A3-FGF2} in this case and hence
\begin{equation}
\Xi_{0,0,1}^{\hat{A}_{3}}\left(\kappa;\bm{\zeta}\right)={\rm Det}\left(1+\kappa\hat{\rho}_{0,0,1}^{\hat{A}_{3}}\left(\hat{x},\hat{p};\bm{\zeta}\right)\right).
\end{equation}
Thus we find an exact relation
\begin{equation}
\frac{\Xi_{4,2}^{{\rm U}\times{\rm U}}\left(\kappa;0\right)}{Z_{4,2}^{{\rm U}}\left(0\right)}=\Xi_{0,0,1}^{\hat{A}_{3}}\left(\kappa;\bm{\zeta}\right).\label{eq:DualityM2-GPF1}
\end{equation}

\subsubsection{From ${\rm O}\left(2N+2\right)_{2}\times{\rm USp}\left(2N\right)_{-1}$ to $\hat{D}_{3}$}

In this section we show the relation between the ${\rm O}\left(2N+2\right)_{2}\times{\rm USp}\left(2N\right)_{-1}$ theory and the $\hat{D}_{3}$ theory.

The ${\rm O}\times{\rm USp}$ density matrix is given by \eqref{eq:DM-OUSpM2}
\begin{equation}
\hat{\rho}_{1,2}^{{\rm O}\times{\rm USp}}\left(\hat{x},\hat{p};0\right)=\frac{\tanh\frac{\hat{x}}{2}}{2\cosh\hat{x}}\left\{ \tanh\frac{\hat{x}}{2},\frac{1}{2\cosh\frac{\hat{p}}{2}}\right\} .
\end{equation}
After performing the canonical transformation as well as the ${\rm O}\left(2N\right)_{2}\times{\rm USp}\left(2N\right)_{-1}$ case
\begin{equation}
\hat{x}\rightarrow-\hat{p},\quad\hat{p}\rightarrow\hat{x},
\end{equation}
we obtain
\begin{equation}
\hat{\rho}_{1,2}^{{\rm O}\times{\rm USp}}\left(\hat{x},\hat{p};0\right)\sim\frac{1}{2\cosh\frac{\hat{p}+\frac{1}{2}\pi i}{2}}\tanh\frac{\hat{p}}{2}\frac{1}{2\cosh\frac{\hat{p}-\frac{1}{2}\pi i}{2}}\left\{ \tanh\frac{\hat{p}}{2},\frac{1}{2\cosh\frac{\hat{x}}{2}}\right\} .
\end{equation}

On the other hand, the $\hat{D}_{3}$ density matrix with the parameter \eqref{eq:DualityM2-para} can be read off from \eqref{eq:DM-D32} as
\begin{equation}
\hat{\rho}_{0,0,1}^{\hat{D}_{3}}\left(\hat{x},\hat{p};\bm{\zeta}\right)=\frac{1}{2}e^{-\frac{1}{4}\hat{x}}\tanh\frac{\hat{p}}{2}e^{-\frac{1}{4}\hat{x}}\frac{1}{2\cosh\frac{\hat{p}}{2}}e^{\frac{1}{4}\hat{x}}\left\{ \frac{1}{2\cosh\frac{\hat{x}}{2}},\tanh\frac{\hat{p}}{2}\right\} e^{\frac{1}{4}\hat{x}}\frac{1}{2\cosh\frac{\hat{p}}{2}}.
\end{equation}
Thus we find the exact relation between the ${\rm O}\left(2N+1\right)_{2}\times{\rm USp}\left(2N\right)_{-1}$ and $\hat{D}_{3}$ matrix models
\begin{equation}
\hat{\rho}_{1,2}^{{\rm O}\times{\rm USp}}\left(\hat{x},\hat{p};0\right)\sim\hat{\rho}_{0,0,1}^{\hat{D}_{3}}\left(\hat{x},\hat{p};\bm{\zeta}\right).\label{eq:DualityM2-DM2}
\end{equation}

Now we move to the grand partition function. The result of the Fermi gas formalism for ${\rm O}\times{\rm USp}$ grand partition function is given by \eqref{eq:GPF-OUSpM2-FGF} in this case and hence
\begin{equation}
\Xi_{1,2}^{{\rm O}\times{\rm USp}}\left(\kappa;0\right)=Z_{2,2}^{{\rm O}}\sqrt{{\rm Det}\left(1+\kappa\hat{\rho}_{1,2}^{{\rm O}\times{\rm USp}}\left(\hat{x},\hat{p};0\right)\right)}.
\end{equation}
The result of the Fermi gas formalism for $\hat{D}_{3}$ grand partition function is given by \eqref{eq:GPF-D3-FGF2} in this case and hence
\begin{equation}
\Xi_{0,0,1}^{\hat{D}_{3}}\left(\kappa;\bm{\zeta}\right)=\sqrt{{\rm Det}\left(1+\kappa\hat{\rho}_{0,0,1}^{\hat{D}_{3}}\left(\hat{x},\hat{p};\bm{\zeta}\right)\right)}.
\end{equation}
Thus we find an exact relation
\begin{equation}
\frac{\Xi_{1,2}^{{\rm O}\times{\rm USp}}\left(\kappa;0\right)}{Z_{2,2}^{{\rm O}}}=\Xi_{0,0,1}^{\hat{D}_{3}}\left(\kappa;\bm{\zeta}\right).\label{eq:DualityM2-GPF2}
\end{equation}

By combining this result and \eqref{eq:DualityM2-GPF1}, we obtain
\begin{equation}
\frac{\Xi_{4,2}^{{\rm U}\times{\rm U}}\left(\kappa;0\right)}{Z_{4,2}^{{\rm U}}\left(0\right)}=\frac{\Xi_{1,2}^{{\rm O}\times{\rm USp}}\left(\kappa;0\right)}{Z_{2,2}^{{\rm O}}}.
\end{equation}
This equality immediately leads to the final result
\begin{equation}
\frac{Z_{4,\left(N+2,N\right)}^{{\rm U}\times{\rm U}}\left(0\right)}{Z_{4,2}^{{\rm U}}\left(0\right)}=\frac{Z_{1,\left(2N+2,2N\right)}^{{\rm O}\times{\rm USp}}\left(0\right)}{Z_{2,2}^{{\rm O}}}.\label{eq:DualityM2-PF}
\end{equation}

At first sight this relation would imply that an additional ${\rm U}\left(2\right)_{4}$ decoupled topological sector (which corresponds to $Z_{4,2}^{{\rm U}}\left(0\right)$) and an additional ${\rm O}\left(2\right)_{2}$ decoupled topological sector (which corresponds to $Z_{2,2}^{{\rm O}}$) are necessary in the right and left side, respectively. However, interestingly, the values of both of the factors are equal since
\begin{equation}
\left|Z_{4,2}^{{\rm U}}\left(0\right)\right|=Z_{2,2}^{{\rm O}}=\frac{1}{2\sqrt{2}}.
\end{equation}
Hence the neither decoupled sector is necessary at least from the viewpoint of the partition function. Therefore, we have rigorously checked the ${\rm U}\left(N+2\right)_{4}\times{\rm U}\left(N\right)_{-4}\Leftrightarrow{\rm O}\left(2N+2\right)_{2}\times{\rm USp}\left(2N\right)_{-1}$ duality by proving the exact equality of the absolute values of the matrix models for arbitrary rank $N$. It would be nice if one can investigate these decoupled sectors from different viewpoint.

\section{Comments on functional relations with chiral projections\label{sec:Functional-Relations}}

In this section we comment on conjectured functional relations between the grand partition functions with and without the chiral projections which were found in \cite{Grassi:2014uua}. They are
\begin{align}
{\rm Det}\left(1+\kappa\hat{\rho}_{4,0}^{{\rm U}\times{\rm U}}\left(\hat{x},\hat{p};0\right)\right) & ={\rm Det}\left(1+\kappa\hat{\rho}_{2,1}^{{\rm U}\times{\rm U}}\left(\hat{x},\hat{p};0\right)\frac{1+\hat{R}}{2}\right),\nonumber \\
{\rm Det}\left(1+\kappa\hat{\rho}_{4,1}^{{\rm U}\times{\rm U}}\left(\hat{x},\hat{p};0\right)\right) & ={\rm Det}\left(1+\kappa\hat{\rho}_{2,0}^{{\rm U}\times{\rm U}}\left(\hat{x},\hat{p};0\right)\frac{1-\hat{R}}{2}\right),\nonumber \\
{\rm Det}\left(1+\kappa\hat{\rho}_{4,2}^{{\rm U}\times{\rm U}}\left(\hat{x},\hat{p};0\right)\right) & ={\rm Det}\left(1+\kappa\hat{\rho}_{2,1}^{{\rm U}\times{\rm U}}\left(\hat{x},\hat{p};0\right)\frac{1-\hat{R}}{2}\right),\nonumber \\
{\rm Det}\left(1+\kappa\hat{\rho}_{8,2}^{{\rm U}\times{\rm U}}\left(\hat{x},\hat{p};0\right)\right) & ={\rm Det}\left(1+\kappa\hat{\rho}_{4,2}^{{\rm U}\times{\rm U}}\left(\hat{x},\hat{p};0\right)\frac{1-\hat{R}}{2}\right).\label{eq:GHMrelation}
\end{align}
These relations have been checked only numerically, and a proof has been missing.

It is also known that the ${\rm O}\times{\rm USp}$ matrix models can be written in terms of the ${\rm U}\times{\rm U}$ density matrix with the chiral projection when there are no mass deformations \cite{Honda:2014ica,Moriyama:2015asx,Moriyama:2016kqi,Moriyama:2016xin,Honda:2015rbb}. In terms of the normalized matrix models, the relations can be written as
\begin{align}
 & \frac{Z_{k,\left(2N,2\left(N+M\right)\right)}^{{\rm O}\times{\rm USp}}\left(0\right)}{Z_{k,\left(0,2M\right)}^{{\rm O}\times{\rm USp}}\left(0\right)}=\int\frac{d^{N}\mu}{N!}\det\left(\left[\braket{\mu_{a}|\hat{\rho}_{2k,2M+1}^{{\rm U}\times{\rm U}}\left(\hat{x},\hat{p};0\right)\frac{1+\hat{R}}{2}|\mu_{b}}\right]_{a,b}^{N\times N}\right),\nonumber \\
 & \frac{Z_{k,\left(2\left(N+M+1\right),2N\right)}^{{\rm O}\times{\rm USp}}\left(0\right)}{Z_{k,\left(2\left(M+1\right),0\right)}^{{\rm O}\times{\rm USp}}\left(0\right)}=\int\frac{d^{N}\mu}{N!}\det\left(\left[\braket{\mu_{a}|\hat{\rho}_{2k,2M+1}^{{\rm U}\times{\rm U}}\left(\hat{x},\hat{p};0\right)\frac{1-\hat{R}}{2}|\mu_{b}}\right]_{a,b}^{N\times N}\right),\nonumber \\
 & \frac{Z_{k,\left(2\left(N+M\right)+1,2N\right)}^{{\rm O}\times{\rm USp}}\left(0\right)}{Z_{k,\left(2M+1,0\right)}^{{\rm O}\times{\rm USp}}\left(0\right)}=\frac{Z_{k,\left(2N+1,2\left(N+M\right)\right)}^{{\rm O}\times{\rm USp}}\left(0\right)}{Z_{k,\left(1,2M\right)}^{{\rm O}\times{\rm USp}}\left(0\right)}\nonumber \\
 & =\int\frac{d^{N}\mu}{N!}\det\left(\left[\braket{\mu_{a}|\hat{\rho}_{2k,2M}^{{\rm U}\times{\rm U}}\left(\hat{x},\hat{p};0\right)\frac{1-\hat{R}}{2}|\mu_{b}}\right]_{a,b}^{N\times N}\right).\label{eq:CPrelation}
\end{align}
Here $M\geq0$. Note that the second line was used in section \ref{subsec:FGF-OUSp-M2}. These relations have been rigorously proven.

By combining \eqref{eq:GHMrelation} and \eqref{eq:CPrelation}, one can obtain the following relations\footnote{This argument for the second line has already been suggested in \cite{Honda:2015rbb}.}
\begin{align}
{\rm Det}\left(1+\kappa\hat{\rho}_{4,0}^{{\rm U}\times{\rm U}}\left(\hat{x},\hat{p};0\right)\right) & ={\rm Det}\left(1+\kappa\hat{\rho}_{1,0}^{{\rm O}\times{\rm USp}}\left(\hat{x},\hat{p};0\right)\right),\nonumber \\
{\rm Det}\left(1+\kappa\hat{\rho}_{4,1}^{{\rm U}\times{\rm U}}\left(\hat{x},\hat{p};0\right)\right) & ={\rm Det}\left(1+\kappa\hat{\rho}_{1,1}^{{\rm O}\times{\rm USp}}\left(\hat{x},\hat{p};0\right)\right),\nonumber \\
{\rm Det}\left(1+\kappa\hat{\rho}_{4,2}^{{\rm U}\times{\rm U}}\left(\hat{x},\hat{p};0\right)\right) & ={\rm Det}\left(1+\kappa\hat{\rho}_{1,2}^{{\rm O}\times{\rm USp}}\left(\hat{x},\hat{p};0\right)\right),\nonumber \\
{\rm Det}\left(1+\kappa\hat{\rho}_{8,2}^{{\rm U}\times{\rm U}}\left(\hat{x},\hat{p};0\right)\right) & ={\rm Det}\left(1+\kappa\hat{\rho}_{2,3}^{{\rm O}\times{\rm USp}}\left(\hat{x},\hat{p};0\right)\right).\label{eq:FunctionalEq}
\end{align}
The first three relations are nothing but the relations we proved, which comes from the dualities between the ${\rm U}\times{\rm U}$ theories and the ${\rm O}\times{\rm USp}$ theories. Therefore, by inverting the argument, we can prove the first three equations of \eqref{eq:GHMrelation}. On the other hand, the fourth relation has not been proven yet, and its physical origin is still mysterious.

\section{Conclusion and discussion\label{sec:Discussion}}

In this paper we studied the dualities between the ABJ(M) theories with ${\rm U}\left(N+M\right)_{4}\times{\rm U}\left(N\right)_{-4}$ and ${\rm O}\left(2N+M\right)_{2}\times{\rm USp}\left(2N\right)_{-1}$ gauge groups ($M=0,1,2$) by focusing on the $S^{3}$ partition function. By using the Fermi gas formalism, we could find the exact relations of the partition functions between the dual theories. We also checked the correspondence of the flavor symmetries between the dual theories for $M=0,1$ cases by introducing the mass deformation and seeing the relation between the deformation parameters. Moreover, we saw that although the exact relations between the partition functions for non-uniform rank theories \eqref{eq:DualityM1-PF} and \eqref{eq:DualityM2-PF} include $N$-independent overall factors, they non-trivially cancel each other, and thus the absolute values of the matrix models for dual theories agree perfectly.

There are various interesting directions for further study. First, in this paper we find the exact relations by using \eqref{eq:UU-A3-D3-OSp}. However, the physical interpretation of this relation is still missing. Here we would like to clarify that we do not claim that the relations ${\rm U}\times{\rm U}\rightarrow\hat{A}_{3}$ and ${\rm O}\times{\rm USp}\rightarrow\hat{D}_{3}$ are the dualities. For example, the FI parameters appeared in this paper for $\hat{A}_{3}$ and $\hat{D}_{3}$ are fixed, which implies the mismatch of the symmetries. Note that the geometry corresponding to $\hat{A}_{3}=\hat{D}_{3}$ theories is known, which is $\mathbb{C}^{2}/\mathbb{Z}_{2}\times\mathbb{C}^{2}/\mathbb{Z}_{2}$ for $M=0,1$ (figure \ref{fig:DualityM0} and \ref{fig:DualityM1}) and $\mathbb{C}^{2}/\mathbb{Z}_{4}\times\mathbb{C}^{2}$ for $M=2$ (figure \ref{fig:DualityM2}). On the other hand, the geometry for the ${\rm U}\times{\rm U}$ and ${\rm O}\times{\rm USp}$ theories is $\mathbb{C}^{4}/\mathbb{Z}_{4}$. Hence the volume of the geometries is the same. Since the volume is the coefficient of the $N^{3/2}$ term of the free energy \cite{Herzog:2010hf}, the equality is the necessary condition of the agreement of the partition functions. From another viewpoint, it was pointed out in \cite{Drukker:2015awa,Assel:2015hsa} that in the Fermi gas formalism, the canonical transformation $\left(\hat{x},\hat{p}\right)\rightarrow\left(-\hat{p},\hat{x}\right)$, which appeared, for example, in \eqref{eq:DualM0-Can1} and \eqref{eq:DualM0-Can2}, implies the mirror symmetry.

Second, it is known that the grand partition functions of the $\hat{A}_{n}$ ($n=2,4$) theories have integrable structures \cite{Bonelli:2017gdk,Nosaka:2020tyv,Bonelli:2022dse,Moriyama:2023mjx}. On the other hand, integrable structures for the ${\rm O}\times{\rm USp}$ theories are not known. As one can see in \eqref{eq:CPrelation}, the grand partition functions of the ${\rm O}\times{\rm USp}$ theories are expressed with the chiral projection. It was found in \cite{Grassi:2014uua} that this type of the Fredholm determinant satisfies the Wronskian-like equations, and it was further pointed out that the Wronskian-like equations leads to the $q$-Painlev\'e ${\rm III}_{3}$ equation \cite{Bonelli:2017gdk,Bershtein:2018zcz}. Integrable structures for the $\hat{D}_{n}$ theories are also not known. Because the Fermi gas formalism played an important role for finding the integrable structure for $\hat{A}_{n}$-type theories, and in this paper we find the exact relations under the Fermi gas formalism, it would be nice if our result helps to find the integrable structures of the ${\rm O}\times{\rm USp}$ and $\hat{D}_{3}$ theories.

Third, it would be interesting whether one can apply the $\hat{A}_{3}=\hat{D}_{3}$ technique used in this paper to other dualities. For example, as we discussed in section \ref{sec:Functional-Relations}, $k=8$ functional relation was found (the last line of \eqref{eq:GHMrelation}), which implies the last line of \eqref{eq:FunctionalEq}. It would be great to apply our technique, or generalized one, to this relation. It would be also applicable for other geometrical dualities, for example dualities coming from $\mathbb{C}^{2}/\mathbb{Z}_{2}\times\mathbb{C}^{2}$ or $\left(\mathbb{C}^{2}/\mathbb{Z}_{2}\times\mathbb{C}^{2}\right)/\mathbb{Z}_{2}$ suggested in \cite{Gang:2011xp}.

Fourth, it would be important to check the dualities with various physical quantities. The localization technique can be applied, for example, for the partition function on a squashed three sphere or lens space \cite{Hama:2011ea,Imamura:2011wg,Imamura:2012rq,Imamura:2013qxa,Kubo:2021xxi}. In addition to the localization, the Fermi gas formalism for the Wilson loop is also known \cite{Klemm:2012ii,Hatsuda:2013yua,Kiyoshige:2016lno}. It would be nice if our derivation can be applied also for these quantities.

\section*{Acknowledgements}

The author thanks Keita Nii, Tomoki Nosaka and Yuya Tanizaki for the collaboration during the early stages of the work. We are also grateful to Yi Pang for valuable discussions. NK is partially supported by National key research and development program under grant No. 2022YFE0134300.

\appendix

\section{Quantum mechanics\label{sec:QM}}

In this appendix we provide notation, definitions and formulas related to the quantum mechanics. In this appendix, we assume that the commutation relation of position operator $\hat{x}$ and momentum operator $\hat{p}$ satisfies $\left[\hat{x},\hat{p}\right]=i\hbar$. A symbol $\ket{\cdot}$ and $\kket{\cdot}$ denote a position eigenvector and a momentum eigenvector, respectively. These vectors are normalized so that the inner products of them satisfy
\begin{align}
 & \braket{x_{1}|x_{2}}=\delta\left(x_{1}-x_{2}\right),\quad\bbrakket{p_{1}|p_{2}}=\delta\left(p_{1}-p_{2}\right),\nonumber \\
 & \brakket{x|p}=\frac{1}{\sqrt{2\pi\hbar}}e^{\frac{i}{\hbar}px},\quad\bbraket{p|x}=\frac{1}{\sqrt{2\pi\hbar}}e^{-\frac{i}{\hbar}px}.\label{eq:Normalization}
\end{align}
The operators satisfy the following formulas for similarity transformations
\begin{align}
 & e^{-\frac{i\eta}{\hbar}\hat{x}}f\left(\hat{p}\right)e^{\frac{i\eta}{\hbar}\hat{x}}=f\left(\hat{p}+\eta\right),\quad e^{-\frac{i\eta}{\hbar}\hat{p}}f\left(\hat{x}\right)e^{\frac{i\eta}{\hbar}\hat{p}}=f\left(\hat{x}-\eta\right),\nonumber \\
 & e^{\mp\frac{i}{2\hbar}\hat{p}^{2}}e^{\mp\frac{i}{2\hbar}\hat{x}^{2}}f\left(\hat{p}\right)e^{\pm\frac{i}{2\hbar}\hat{x}^{2}}e^{\pm\frac{i}{2\hbar}\hat{p}^{2}}=f\left(\pm\hat{x}\right).\label{eq:OpSim}
\end{align}
Correspondingly, the vectors satisfy the following formulas
\begin{align}
 & e^{\frac{i\eta}{\hbar}\hat{x}}\kket p=\kket{p+\eta},\quad\bbra pe^{-\frac{i\eta}{\hbar}\hat{x}}=\bbra{p+\eta},\nonumber \\
 & e^{\mp\frac{i}{2\hbar}\hat{p}^{2}}e^{\mp\frac{i}{2\hbar}\hat{x}^{2}}\kket p=i^{\mp\frac{1}{2}}e^{\pm\frac{i}{2\hbar}p^{2}}\ket{\pm p},\quad\bbra pe^{\pm\frac{i}{2\hbar}\hat{x}^{2}}e^{\pm\frac{i}{2\hbar}\hat{p}^{2}}=i^{\pm\frac{1}{2}}e^{\mp\frac{i}{2\hbar}p^{2}}\bra{\pm p}.\label{eq:VecSim}
\end{align}
The operators also satisfy the Campbell-Baker-Hausdorff formula
\begin{equation}
e^{c_{1}\hat{x}}e^{c_{2}\hat{p}}=e^{\frac{1}{2}ic_{1}c_{2}\hbar}e^{c_{1}\hat{x}+c_{2}\hat{p}}.
\end{equation}

In this paper the reflection operator plays an important role. The reflection operator is defined by
\begin{equation}
\hat{R}=\int dx\ket x\bra{-x}.
\end{equation}
This operator satisfies
\begin{equation}
\hat{R}^{2}=1,\quad\hat{x}\hat{R}=-\hat{R}\hat{x},\quad\hat{p}\hat{R}=-\hat{R}\hat{p}.
\end{equation}

We define the transposed operator $\hat{A}^{t}$ by
\begin{equation}
\braket{x|\hat{A}^{t}|y}=\braket{y|\hat{A}|x}.
\end{equation}
It satisfies
\begin{align}
\left(\hat{A}\hat{B}\right)^{t} & =\hat{B}^{t}\hat{A}^{t},\quad\left(c\hat{A}\right)^{t}=c\hat{A}^{t},\quad\left(\hat{A}^{-1}\right)^{t}=\left(\hat{A}^{t}\right)^{-1},\nonumber \\
f\left(\hat{x}\right)^{t} & =f\left(\hat{x}\right),\quad f\left(\hat{p}\right)^{t}=f\left(-\hat{p}\right),\quad\hat{R}^{t}=\hat{R}.
\end{align}

\section{Formulas for Fermi gas formalism\label{sec:Formulas}}

In this appendix we enumerate formulas which are used for applying the Fermi gas formalism.

\subsection{Determinant formula\label{subsec:CauchyDet}}

In this section we provide formulas which can be apply to the 1-loop determinant factor. In this section we use the quantum mechanics introduced in appendix \ref{sec:QM} with $\hbar=2\pi\ell$.

The 1-loop determinant factors can be written in terms of an appropriate matrix as
\begin{align}
 & \frac{\prod_{a<a'}^{N}2\sinh\frac{\alpha_{a}-\alpha_{a'}}{2\ell}\prod_{b<b'}^{N+M}2\sinh\frac{\beta_{b}-\beta_{b'}}{2\ell}}{\prod_{a}^{N}\prod_{b}^{N+M}2\cosh\frac{\alpha_{a}-\beta_{b}}{2\ell}}=\det\left(\begin{array}{c}
\left[\hbar\braket{\alpha_{a}|\frac{1}{2\cosh\frac{\hat{p}-i\pi M}{2}}|\beta_{b}}\right]_{a,b}^{N\times\left(N+M\right)}\\
\left[\frac{\hbar}{\sqrt{\ell}}\bbraket{t_{M,j}|\beta_{b}}\right]_{j,b}^{M\times\left(N+M\right)}
\end{array}\right),\nonumber \\
 & \frac{\prod_{a<a'}^{N+M}2\sinh\frac{\alpha_{a}-\alpha_{a'}}{2\ell}\prod_{b<b'}^{N}2\sinh\frac{\beta_{b}-\beta_{b'}}{2\ell}}{\prod_{a}^{N+M}\prod_{b}^{N}2\cosh\frac{\alpha_{a}-\beta_{b}}{2\ell}}\nonumber \\
 & =\det\left(\begin{array}{cc}
\left[\hbar\braket{\alpha_{a}|\frac{1}{2\cosh\frac{\hat{p}+i\pi M}{2}}|\beta_{b}}\right]_{a,b}^{\left(N+M\right)\times N} & \left[\frac{\hbar}{\sqrt{\ell}}\brakket{\alpha_{a}|-t_{M,j}}\right]_{a,j}^{\left(N+M\right)\times M}\end{array}\right),\label{eq:DetFormula-cosh}
\end{align}
where
\begin{equation}
t_{M,j}=2\pi i\left(\frac{M+1}{2}-j\right).\label{eq:t-Def}
\end{equation}
Similarly,
\begin{align}
 & \frac{\prod_{a<a'}^{N+M}2\sinh\frac{\alpha_{a}-\alpha_{a'}}{2\ell}\prod_{b<b'}^{N}2\sinh\frac{\beta_{b}-\beta_{b'}}{2\ell}}{\prod_{a}^{N+M}\prod_{b}^{N}2\sinh\frac{\alpha_{a}-\beta_{b}}{2\ell}}\nonumber \\
 & =\left(-1\right)^{MN}i^{-N^{2}}\det\left(\begin{array}{cc}
\left[\hbar\braket{\alpha_{a}|\frac{1}{2}\tanh\frac{\hat{p}-i\pi M}{2}|\beta_{b}}\right]_{a,b}^{\left(N+M\right)\times N} & \left[\frac{\hbar}{\sqrt{\ell}}\brakket{\alpha_{a}|-t_{M,j}}\right]_{a,j}^{\left(N+M\right)\times M}\end{array}\right).\label{eq:DetFormula-sinh}
\end{align}

These formulas follows from the following two formulas. First formula is the Cauchy-Vandermonde determinant formula \cite{Matsumoto:2013nya}
\begin{align}
 & \frac{\prod_{a<a'}^{N+M}2\sinh\frac{\alpha_{a}-\alpha_{a'}}{2\ell}\prod_{b<b'}^{N}2\sinh\frac{\beta_{b}-\beta_{b'}}{2\ell}}{\prod_{a}^{N+M}\prod_{b}^{N}2\cosh\frac{\alpha_{a}-\beta_{b}}{2\ell}}\nonumber \\
 & =\det\left(\begin{array}{cc}
\left[\left(-1\right)^{M}\frac{e^{-\frac{M}{2\ell}\left(\alpha_{a}-\beta_{b}\right)}}{2\cosh\frac{\alpha_{a}-\beta_{b}}{2\ell}}\right]_{a,b}^{\left(N+M\right)\times N} & \left[e^{\frac{1}{\ell}\alpha_{a}t_{M,j}}\right]_{a,j}^{\left(N+M\right)\times M}\end{array}\right).\label{eq:CVDetFormula}
\end{align}
By shifting the variables as $\alpha\rightarrow\alpha-\pi i\ell$, we also obtain
\begin{align}
 & \frac{\prod_{a<a'}^{N+M}2\sinh\frac{\alpha_{a}-\alpha_{a'}}{2\ell}\prod_{b<b'}^{N}2\sinh\frac{\beta_{b}-\beta_{b'}}{2\ell}}{\prod_{a}^{N+M}\prod_{b}^{N}2\sinh\frac{\alpha_{a}-\beta_{b}}{2\ell}}\nonumber \\
 & =\left(-1\right)^{MN}i^{-N^{2}}\det\left(\begin{array}{cc}
\left[i\frac{e^{-\frac{M}{2\ell}\left(\alpha_{a}-\beta_{b}\right)}}{2\sinh\frac{\alpha_{a}-\beta_{b}}{2\ell}}\right]_{a,b}^{\left(N+M\right)\times N} & \left[e^{\frac{1}{\ell}\alpha_{a}t_{M,j}}\right]_{a,j}^{\left(N+M\right)\times M}\end{array}\right).\label{eq:CVDetFormula2}
\end{align}
Second formulas are the Fourier transformation formulas
\begin{align}
\frac{1}{2\cosh\frac{\alpha-\beta}{2\ell}} & =\frac{1}{2\pi}\int dp\frac{e^{\frac{i}{2\pi\ell}p\left(\alpha-\beta\right)}}{2\cosh\frac{p}{2}}=\hbar\braket{\alpha|\frac{1}{2\cosh\frac{\hat{p}}{2}}|\beta},\nonumber \\
\frac{i}{2\sinh\frac{\alpha-\beta}{2\ell}} & =\frac{1}{2\pi}\int dpe^{\frac{i}{2\pi\ell}p\left(\alpha-\beta\right)}\frac{1}{2}\tanh\frac{p}{2}=\frac{\hbar}{2}\braket{\alpha|\tanh\frac{\hat{p}}{2}|\beta}.\label{eq:FourierFormula}
\end{align}
The combination of \eqref{eq:CVDetFormula} (and its transpose) and the first line of \eqref{eq:FourierFormula} leads to \eqref{eq:DetFormula-cosh}, while the combination of \eqref{eq:CVDetFormula2} and the second line of \eqref{eq:FourierFormula} leads to \eqref{eq:DetFormula-cosh}. Here we also need to use the formulas \eqref{eq:OpSim} and \eqref{eq:VecSim} appropriately. Note that for obtaining the determinant formula \eqref{eq:DetFormula-cosh}, we need to change the integration counter to the imaginary direction. See also appendix A.1 of \cite{Kubo:2020qed} for this point.

\subsection{Fredholm determinant formula\label{subsec:FredholmDet}}

In this section we show three types of Fredholm determinant formulas. The first two formulas appear when we move from the canonical partition function to the grand partition function
\begin{equation}
\Xi^{\left(i\right)}\left(\kappa\right)=\sum_{N=0}^{\infty}\kappa^{N}\hat{Z}_{N}^{\left(i\right)},
\end{equation}
where we assume that $\hat{Z}^{\left(i\right)}\left(N\right)$ is normalized so that $\hat{Z}^{\left(i\right)}\left(0\right)=1$. The formulas are listed in this section, and their proofs are in the subsections.

Note that in the main text, the partition function is sometimes not normalized. However, even in that case, if the partition function is factorized into an $N$-independent part and an $N$-dependent part which is normalized so that it is $1$ when $N=0$, and in this paper it is always the case, this formula can be applied. The resulting grand partition function is multiplied by the $N$-independent part.

The first formula is for
\begin{equation}
\hat{Z}_{N}^{\left(1\right)}=\int\frac{d^{N}\mu}{N!}\det\left(\left[\braket{\mu_{a}|\hat{\rho}_{1}|\mu_{b}}\right]_{a,b}^{N\times N}\right).
\end{equation}
This type of partition function appears in the circular quiver theory where all nodes are unitary group. The grand partition function is given by \cite{Marino:2011eh}
\begin{equation}
\Xi^{\left(1\right)}\left(\kappa\right)={\rm Det}\left(1+\kappa\hat{\rho}_{1}\right),\label{eq:GPF-Circle}
\end{equation}
where ${\rm Det}$ is a Fredholm determinant.

The second formula is for
\begin{align}
\hat{Z}_{N}^{\left(2\right)} & =\frac{1}{2^{2N}}\int\frac{d^{N}\mu}{N!}\frac{d^{N}\mu'}{N!}\frac{d^{N}\nu}{N!}\frac{d^{N}\nu'}{N!}\nonumber \\
 & \quad\times\det\left(\left[\braket{\mu_{a}|\hat{\rho}_{{\rm L}}|\mu_{b}'}\right]_{a,b}^{N\times N}\right)\det\left(\left[\braket{\bar{\mu}_{a}|\hat{\rho}_{{\rm C}}|\bar{\nu}_{b}}\right]_{a,b}^{2N\times2N}\right)\det\left(\left[\braket{\nu_{a}'|\hat{\rho}_{{\rm R}}|\nu_{b}}\right]_{a,b}^{N\times N}\right),
\end{align}
where
\begin{equation}
\bar{\mu}_{a}=\begin{cases}
\mu_{a} & \left(1\leq a\leq N\right)\\
\mu_{a-N}' & \left(N+1\leq a\leq2N\right)
\end{cases},\quad\bar{\nu}_{a}=\begin{cases}
\nu_{a} & \left(1\leq a\leq N\right)\\
\nu_{a-N}' & \left(N+1\leq a\leq2N\right)
\end{cases}.
\end{equation}
This type of partition function appears in the orientifold theory or the $\hat{D}$-type quiver theory where all nodes are unitary group. The grand canonical expression of this partition function becomes
\begin{align}
\Xi^{\left(2\right)}\left(\kappa\right) & =\sqrt{{\rm Det}\left(1+\kappa\hat{\rho}_{2}\right)},\label{eq:GPF-Orient}
\end{align}
where
\begin{equation}
\hat{\rho}_{2}=\frac{1}{4}\left(\hat{\rho}_{{\rm L}}-\hat{\rho}_{{\rm L}}^{t}\right)\hat{\rho}_{{\rm C}}\left(\hat{\rho}_{{\rm R}}-\hat{\rho}_{{\rm R}}^{t}\right)\hat{\rho}_{{\rm C}}^{t}.\label{eq:rho-Orient}
\end{equation}

The third formula is for the Fredholm determinant with the chiral projection
\begin{equation}
\Xi^{\left(3\right)}\left(\kappa\right)={\rm Det}\left(1+\kappa\hat{\rho}_{3}\frac{1\pm\hat{R}}{2}\right).
\end{equation}
This type of grand partition function appears also in the orientifold theory or the $\hat{D}$-type quiver theory. If the density matrix satisfies
\begin{equation}
{\rm tr}\left[\hat{\rho}_{3}^{n}\hat{R}\right]=0,\quad\left[\hat{\rho}_{3},\hat{R}\right]=0,\label{eq:GPF-DMcond}
\end{equation}
for any positive integer $n$, then
\begin{align}
\Xi^{\left(3\right)}\left(\kappa\right) & =\sqrt{{\rm Det}\left(1+\kappa\hat{\rho}_{3}\right)}.\label{eq:GPF-CP}
\end{align}
This expression is the same with \eqref{eq:GPF-Orient}.

\subsubsection{Proof for second formula}

In this section we show how we obtain the second formula \eqref{eq:GPF-Orient}.

First, after using the diagonalization formula \eqref{eq:DiagFormula} to the first and third determinant, we can put them into the second determinant and obtain
\begin{equation}
\hat{Z}_{N}^{\left(2\right)}=\frac{1}{2^{2N}}\int\frac{d^{N}\mu}{N!}\frac{d^{N}\nu}{N!}\det\left(\begin{array}{cc}
\left[\braket{\mu_{a}|\hat{\rho}_{{\rm C}}|\nu_{b}}\right]_{a,b}^{N\times N} & \left[\braket{\mu_{a}|\hat{\rho}_{{\rm C}}\hat{\rho}_{{\rm R}}|\nu_{b}}\right]_{a,b}^{N\times N}\\
\left[\braket{\mu_{a}|\hat{\rho}_{{\rm L}}\hat{\rho}_{{\rm C}}|\nu_{b}}\right]_{a,b}^{N\times N} & \left[\braket{\mu_{a}|\hat{\rho}_{{\rm L}}\hat{\rho}_{{\rm C}}\hat{\rho}_{{\rm R}}|\nu_{b}}\right]_{a,b}^{N\times N}
\end{array}\right).
\end{equation}
For proceeding, we use two formulas. The first formula is a formula which relates this type of integration to a Pfaffian \cite{Moriyama:2015jsa}
\begin{align}
 & \frac{1}{2^{2N}}\int\frac{d^{N}\alpha}{N!}\det\left(\begin{array}{cc}
\left[\phi_{a}\left(\alpha_{b}\right)\right]_{a,b}^{2N\times N} & \left[\psi_{a}\left(\alpha_{b}\right)\right]_{a,b}^{2N\times N}\end{array}\right)\nonumber \\
 & =\left(-1\right)^{\frac{1}{2}N\left(N-1\right)}{\rm Pf}\left(\left[\frac{1}{4}\int d\alpha\left(\phi_{a}\left(\alpha\right)\psi_{b}\left(\alpha\right)-\psi_{a}\left(\alpha\right)\phi_{b}\left(\alpha\right)\right)\right]_{a,b}^{2N\times2N}\right).
\end{align}
The second formula is a formula for a Pfaffian of a matrix whose components satisfy 
\begin{equation}
\braket{\alpha|\hat{{\cal O}}_{ij}|\beta}=-\braket{\beta|\hat{{\cal O}}_{ji}|\alpha}.
\end{equation}
For this matrix, the following formula holds \cite{Moriyama:2015jsa}
\begin{align}
 & \sum_{N=0}^{\infty}\kappa^{N}\left(-1\right)^{\frac{1}{2}N\left(N-1\right)}\int\frac{d^{N}\alpha}{N!}{\rm Pf}\left(\begin{array}{cc}
\left[\braket{\alpha_{a}|\hat{{\cal O}}_{11}|\alpha_{b}}\right]_{a,b}^{N\times N} & \left[\braket{\alpha_{a}|\hat{{\cal O}}_{12}|\alpha_{b}}\right]_{a,b}^{N\times N}\\
\left[\braket{\alpha_{a}|\hat{{\cal O}}_{21}|\alpha_{b}}\right]_{a,b}^{N\times N} & \left[\braket{\alpha_{a}|\hat{{\cal O}}_{22}|\alpha_{b}}\right]_{a,b}^{N\times N}
\end{array}\right)\nonumber \\
 & =\sqrt{{\rm Det}\left(\begin{array}{cc}
1-\hat{{\cal O}}_{21} & \hat{{\cal O}}_{22}\\
\hat{{\cal O}}_{11} & 1+\hat{{\cal O}}_{12}
\end{array}\right)}.
\end{align}
By combining these two formulas, we obtain
\begin{equation}
\Xi^{\left(2\right)}\left(\kappa\right)=\sqrt{{\rm Det}\left[\left(\begin{array}{cc}
1 & \hat{\rho}_{{\rm L}}\\
0 & 1
\end{array}\right)^{-1}\left(\begin{array}{cc}
1 & \hat{\rho}_{{\rm L}}\\
0 & 1+z\hat{\rho}_{{\rm C}}\left(\hat{\rho}_{{\rm R}}^{t}-\hat{\rho}_{{\rm R}}\right)\hat{\rho}_{{\rm C}}^{t}\left(\hat{\rho}_{{\rm L}}^{t}-\hat{\rho}_{{\rm L}}\right)
\end{array}\right)\left(\begin{array}{cc}
1 & -\hat{\rho}_{{\rm L}}\\
0 & 1
\end{array}\right)^{-1}\right]}.
\end{equation}
Since the determinant of the first and the third matrices are $1$, \eqref{eq:GPF-Orient} holds.

\subsubsection{Proof for third formula}

In this section we show how we obtain the third formula \eqref{eq:GPF-CP}. The similar proof has been already appeared in \cite{Assel:2015hsa}.

The main strategy of the proof is to show that the spectrum of the Fredholm determinant of the density matrix with chiral projections $\left(1\pm\hat{R}\right)/2$ are same. The conditions \eqref{eq:GPF-DMcond} are the sufficient condition for this.

By using to the first condition ${\rm tr}\left[\hat{\rho}_{3}^{n}\hat{R}\right]=0$, one can show that
\begin{equation}
{\rm tr}\left[\hat{\rho}_{3}^{n}\frac{1+\hat{R}}{2}\right]={\rm tr}\left[\hat{\rho}_{3}^{n}\frac{1-\hat{R}}{2}\right].
\end{equation}
This equality and the second condition $\left[\hat{\rho}_{3},\hat{R}\right]=0$ leads to
\begin{equation}
{\rm tr}\left[\left(\hat{\rho}_{3}\frac{1+\hat{R}}{2}\right)^{n}\right]={\rm tr}\left[\left(\hat{\rho}_{3}\frac{1-\hat{R}}{2}\right)^{n}\right].
\end{equation}
Since the Fredholm determinant can be expressed by traces as
\begin{equation}
{\rm Det}\left(1+\hat{{\cal O}}\right)=\exp\left[\sum_{\ell=1}^{\infty}\frac{\left(-1\right)^{\ell-1}}{\ell}{\rm tr}\hat{{\cal O}}^{\ell}\right],
\end{equation}
we obtain
\begin{equation}
{\rm Det}\left(1+\kappa\hat{\rho}_{3}\frac{1+\hat{R}}{2}\right)={\rm Det}\left(1+\kappa\hat{\rho}_{3}\frac{1-\hat{R}}{2}\right).
\end{equation}
Since
\begin{equation}
{\rm Det}\left(1+\kappa\hat{\rho}_{3}\frac{1+\hat{R}}{2}\right){\rm Det}\left(1+\kappa\hat{\rho}_{3}\frac{1-\hat{R}}{2}\right)={\rm Det}\left(1+\kappa\hat{\rho}_{3}\right),
\end{equation}
\eqref{eq:GPF-CP} holds.

\printbibliography

\end{document}